\documentclass{aa}

\usepackage{graphicx}
\usepackage{amsmath}
\usepackage[english]{babel}
\usepackage{subfig}
\usepackage{amssymb}
\usepackage{wrapfig}
\usepackage{pdflscape}
\usepackage{lscape}
\usepackage{longtable}
\usepackage{appendix}
\usepackage{textcomp}
\usepackage{multirow}
\bibpunct{(}{)}{;}{a}{}{,}
\usepackage{array}
\usepackage{txfonts}

\begin{document} 

   \title{Large-scale environments of narrow-line Seyfert 1 galaxies}


   \author{E. J\"{a}rvel\"{a}\inst{1}\fnmsep\inst{2}\thanks{\email{emilia.jarvela@aalto.fi}}
          \and 
          A. L\"{a}hteenm\"{a}ki\inst{1}\fnmsep\inst{2}\fnmsep\inst{3}
          \and
          H. Lietzen\inst{3}
          \and
          A. Poudel\inst{4}
          \and
          P. Hein\"{a}m\"{a}ki\inst{4}
          \and
          M. Einasto\inst{3}
          }

   \institute{Aalto University Mets\"{a}hovi Radio Observatory, Mets\"{a}hovintie 114, FI-02540 Kylm\"{a}l\"{a}, Finland
         \and
             Aalto University Department of Electronics and Nanoengineering, P.O. Box 15500, FI-00076 AALTO, Finland
        \and
             Tartu Observatory, Observatooriumi 1, 61602 T\~{o}ravere, Estonia
        \and
             Tuorla Observatory, University of Turku, Väisäläntie 20, Piikkiö, Finland
             }

   \date{Received ; accepted }


  \abstract
   {Studying large-scale environments of narrow-line Seyfert 1 (NLS1) galaxies gives a new perspective on their properties, particularly their radio loudness. The large-scale environment is believed to have an impact on the evolution and intrinsic properties of galaxies, however, NLS1 sources have not been studied in this context before. We have a large and diverse sample of 1341 NLS1 galaxies and three separate environment data sets constructed using Sloan Digital Sky Survey. We use various statistical methods to investigate how the properties of NLS1 galaxies are connected to the large-scale environment, and compare the large-scale environments of NLS1 galaxies with other active galactic nuclei (AGN) classes, for example, other jetted AGN and broad-line Seyfert 1 (BLS1) galaxies, to study how they are related. NLS1 galaxies reside in less dense environments than any of the comparison samples, thus confirming their young age. The average large-scale environment density and environmental distribution of NLS1 sources is clearly different compared to BLS1 galaxies, thus it is improbable that they could be the parent population of NLS1 galaxies and unified by orientation. Within the NLS1 class there is a trend of increasing radio loudness with increasing large-scale environment density, indicating that the large-scale environment affects their intrinsic properties. Our results suggest that the NLS1 class of sources is not homogeneous, and furthermore, that a considerable fraction of them are misclassified. We further support a published proposal to replace the traditional classification to radio-loud, and radio-quiet or radio-silent sources with a division into jetted and non-jetted sources.}

   \keywords{galaxies: active -- galaxies: Seyfert -- large-scale structure of Universe}

   \maketitle


\section{Introduction}
\label{intro}

Active galactic nuclei (AGN) are the most energetic non-transient phenomena in our Universe. They radiate over the whole electromagnetic spectrum, from radio to TeV energies, and are powered by a supermassive black hole accreting matter. AGN can be divided into various classes based on their observed properties. The most numerous classes are the luminous and usually more distant quasars, bright enough to easily overshine their host galaxy, and fainter Seyfert galaxies with often detectable host galaxies. Seyfert galaxies were originally divided into two subclasses based on their optical spectra \citep{1974khachikian1}; Seyfert 1 galaxies (Sy1) show broad permitted emission lines arising
from the broad-line region (BLR) close to the nucleus and narrow forbidden emission lines originating from the narrow-line region (NLR) further away. In Seyfert 2 galaxies (Sy2) the broad emission lines are not directly detectable and they show only the narrow lines in their spectra. It was early on suggested that Sy1 and Sy2 galaxies are similar sources seen at different angles and through varying amounts of obscuring matter \citep{1978osterbrock1}. Soon after, several intermediate types, for example, Sy 1.2 and 1.8 -- corresponding to different viewing angles -- were introduced supporting this scenario \citep{1981osterbrock1}.

A subclass of Sy1 galaxies with narrow permitted lines arising from the BLR was described by \citet{1985osterbrock1}. These narrow-line Seyfert 1 (NLS1) galaxies are defined by their optical spectra; both, permitted and forbidden emission lines are narrow 
\citep[$FWHM$(H$\beta$)$ < 2000$ km s$^{-1}$,][]{1989goodrich1} and [O III] is relatively weak 
\citep[F($\lbrack$O III$\rbrack$)/F(H$\beta) <$ 3,][]{1985osterbrock1}. In addition relative Fe II emission is strong in the majority of NLS1 sources \citep{1985osterbrock1}.

The black hole masses in NLS1 galaxies are low or intermediate \citep[$M_{\text{BH}} < 10^{8}$ $M_{\sun}$,][]{2000peterson1}. They accrete at extraordinarily high rates, from 0.1 Eddington ratio to super-Eddington accretion \citep{1992boroson1}, and lie below the normal relations of the black hole mass and the properties of the galactic bulge -- $\sigma_{\ast}$ (stellar velocity dispersion) and $L_{\text{bulge}}$ \citep[\citet{2001mathur1}, but some studies also disagree,for example,][]{2015woo1}. It has been suggested that NLS1 galaxies are young objects in the early stages of their evolution \citep{2001mathur1}.

The majority of NLS1 sources are radio-quiet; only $\sim$7$\%$ are radio-loud and 2\% -- 3\% percent very radio-loud \citep{2006komossa1}. NLS1 galaxies typically show a very compact unresolved radio core, however, lately, evidence of pc- and kpc-scale structures has been found \citep{2010gliozzi1, 2012doi1, 2015richards1, 2015gu1, 2013doi1, 2015doi1, 2015richards2, 2016lister1}. Subluminal and superluminal speeds measured in some sources suggest Lorentz factors and viewing angles similar to BL Lac objects (BLO) and flat-spectrum radio quasars (FSRQ) \citep{2016lister1}. The \textit{Fermi Gamma-ray Space Telescope} has detected gamma-ray emission in some NLS1 galaxies \citep[{e.g.}][]{2009abdo2}, thus confirming the presence of powerful relativistic jets. NLS1 galaxies detected in gamma-rays or showing extended radio emission seem to have, on average, more massive black holes \citep[$M_{\text{BH}} > 10^7 M_{\sun}$, e.g. ][]{2012doi1,2015foschini1} than the NLS1 population in general \citep[e.g. this paper and ][]{2015jarvela1}. However, these samples are not complete; more studies will be needed to clarify this possible connection.

So far, host galaxy studies of NLS1 galaxies have mostly concentrated on the radio-quiet sources. They have been found to preferably, but not exclusively, be hosted by disk-like galaxies. Only a few studies of the host galaxies of radio-loud NLS1 sources exist; gamma-ray emitting 1H0323+342 resides in a one-armed spiral galaxy \citep{2007zhou1} or possibly in a system disturbed by merging \citep{2008anton1,2014leontavares1}, and the host galaxy of another gamma-ray emitting NLS1, FBQS J1644+2619, seems to be a barred lenticular galaxy \citep{2017olguiniglesias1}. \citet{2000mathur1} suggested that all NLS1 galaxies could be
sources rejuvenated by a recent merger. Later studies \citep[e.g. ][]{2007ohta1,2007ryan1} found no evidence for abundant interaction or merging in NLS1 galaxies, thus it seems improbable that all NLS1 galaxies would be a result of a merger or interaction. However, they might play a role; mergers and interaction are known to be able to trigger the nuclear activity \citep[e.g. ][]{2008barth1,2011ellison1}.

Studying NLS1 galaxies is challenging because of the seemingly heterogeneous nature of the population. One additional way of addressing the issue is to study the environments of NLS1 galaxies compared to other AGN classes and within the NLS1 population. The environment -- at all scales -- affects the intrinsic properties and evolution of galaxies, and consequently their nuclear activity. We can divide the environment into several scales: 1) the host galaxy, which is the closest environment of an AGN; 2) the local environment, including the neighbouring galaxies and the group and cluster the galaxy belongs to ; and 3) the large-scale environment, tracing the largest, supercluster-scale, structures in the Universe.

The connection between the AGN and its host galaxy has been extensively studied \citep[e.g.][]{2008storchi-bergmann1,2010vandeven1,2012povic1,2012fabian1,2015king1}. The host galaxy regulates the gas supply of the black hole and thus has a direct impact on its activity level. The AGN in turn affects the host galaxy via various feedback mechanisms, for example, the radiation pressure, the jet, and winds and outflows. The AGN can induce both negative and positive feedback, and thus steer the evolution of the galaxy by, for example, quenching or enhancing star formation.

The local environment can have an impact on galaxy dynamics and evolution. In a galaxy -- cluster interaction the galaxy might lose a fraction or almost all of its gas due to ram-pressure stripping, thus quenching star formation and advancing the evolution \citep{2014ebeling1,2016steinhauser1}. In a dense enough environment, a galaxy might undergo a number of minor and major mergers that distort its morphology. Mergers may cause gas infall, triggering circumnuclear star formation and feeding the black hole, and replenish or strip gas reservoirs. The connection between the mergers and triggering an AGN is not clear; some studies suggest that both minor \citep{1999taniguchi1,2008barth1} and major \citep{2008urrutia1,2011ellison1} mergers are able to trigger the nuclear activity, whereas others do not find them connected \citep{2000corbin1,2011cisternas1,2012kocevski1}. It has been argued that mergers, especially major ones, predominantly trigger the most luminous AGN \citep{2012treister1,2014villforth1}. 

The connection between cluster-scale environment and galaxy morphology was discovered early on \citep{1931hubble1,1980dressler1} and has been established by subsequent studies \citep[e.g.][]{2009park1,2017chen1}. This relation is thought to be a consequence of the more frequent galaxy -- galaxy interactions in denser regions. Mergers and interaction transform the galaxy morphology towards the early type, and the speed of this evolution depends on the environment density.

The position of the galaxy in the cosmic web of superclusters, filaments and voids affects its properties as well. The effect is similar to the one found for cluster-scale environments; the galaxies residing in denser large-scale environments -- close to or in filaments and superclusters -- are preferably ellipticals, whereas the fraction of spirals increases the further one departs from these higher-density regions \citep{2012lietzen1,2014einasto1,2017chen1,2017kuutma1,2017pandey1}. Moreover, the properties of the galaxy groups are affected by the large-scale environment \citep{2017poudel1}.

In this paper we study the large-scale environments of NLS1 galaxies using diverse and large, statistically significant samples. We use three environment data sets at three different redshift ranges between 0 and 0.62 to compare the large-scale environments of NLS1 sources with the large-scale environments of other types of AGN, as well as study the differences in the environments within the NLS1 population. In addition, we are interested in seeing whether the large-scale environment is connected to any of the intrinsic properties of NLS1 galaxies. 

This paper is organised as follows. In Section~\ref{sec:data} we introduce the sample and the data used for the analyses, and calculate some additional parameters. We present the results in Section~\ref{sec:results}, discuss the main findings in Section~\ref{sec:disc}, and finally conclude in Section~\ref{sec:concl}. Throughout the paper we assume a cosmology with H$_{0}$ = 73 km s$^{-1}$ Mpc$^{-1}$, $\Omega_{\text{matter}}$ = 0.27 and $\Omega_{\text{vacuum}}$ = 0.73 \citep{2007spergel1}.

\section{Data}
\label{sec:data}

\subsection{Sample}

Our seed sample included all 2011 NLS1 galaxies from \citet{2006zhou1} and 38 additional sources chosen from \citet{2006komossa1, 2006whalen1, 2008yuan1} and \citet{2011foschini1} based on radio loudness. These additional sources were included because they are part of the Mets\"{a}hovi NLS1 monitoring programme \citep{2017lahteenmaki1}. NLS1 galaxies in \citet{2006zhou1} were selected from Sloan Digital Sky Survey{\footnote{www.sdss.org}} \citep[SDSS,][]{2000york1} Data Release 3 and had $z \lesssim$ 0.8 
and the FWHM of the broad H$\beta$ or H$\alpha$  $\lesssim$ 2200 km s$^{-1}$ at the 10 $\sigma$ or higher confidence level. Most of these sources have not been detected at radio frequencies. \citet{2008yuan1} presents a sample of 23 radio-loud NLS1 galaxies selected from SDSS Data Release 5 under the same restrictions as in \citet{2006zhou1}. The sample in \citet{2006komossa1} included 11 radio-loud NLS1 galaxies found by cross-correlating the Catalogue of Quasars and Active Nuclei \citep{2003veroncetty1} with several radio and optical catalogs using the cross-matcher application developed within the German Astrophysical Virtual Observatory (GAVO){\footnote{www.g-vo.org}} project. The sample was limited by the requirement H$\beta <$ 2000 km s$^{-1}$. \citet{2011foschini1} and \citet{2006whalen1} both present a sample of both radio-loud and radio-quiet NLS1 galaxies.

\subsection{Multiwavelength and spectral data}

We retrieved all the multiwavelength data available from ASI Science Data Center (ASDC{\footnote{www.asdc.asi.it}}). Data obtained from ASDC were already corrected for galactic extinction. The archival data are not simultaneous. Studying correlations between wavebands of variable sources, such as AGN, should ideally be performed with data that are no more than a couple of weeks apart or even less for more rapidly varying jetted sources. The time delays between the wavebands should also be taken into account, but this would require frequent multiwavelength monitoring of the sources and is, in practise, often impossible. Details of the wavebands used are given below. The data in all wavebands are not complete.

{\bf Radio data} (radio flux density, $F_{\text{R}}$) are from the National Radio Astronomy Observatory (NRAO) Very Large Array (VLA{\footnote{http://www.vla.nrao.edu/}}) Faint Images of the Radio Sky at Twenty-Centimeters (FIRST{\footnote{www.sundog.stsci.edu}}) survey and NRAO VLA Sky Survey (NVSS{\footnote{http://www.cv.nrao.edu/nvss/}}), both at 1.4~GHz. 
In the case of multiple detections we chose the one closest to the source coordinates. The search radius used for the radio surveys was 1 arcmin, so it is possible that 
in cases when the NLS1 source was not detected in radio, the radio data are actually that of a nearby radio source, but the number of the false sources is very low. The detection limit of FIRST survey is 1~mJy.

{\bf Infrared data} (infrared flux density, $F_{\text{IR}}$) are from Wide-field Infrared Survey Explorer (WISE{\footnote{www.nasa.gov/wise}}) AllWISE Data Release{\footnote{http://wise2.ipac.caltech.edu/docs/release/allwise/}}. WISE bands are W1--W4 with wavelengths 3.4, 4.6, 12, and 22 $\mu$m, respectively.

{\bf Optical data} (optical flux density, $F_{\text{O}}$) are from SDSS Data Release 10{\footnote{https://www.sdss3.org/dr10/}} (DR10), or, if DR10 data were not available, from DR7{\footnote{http://www.sdss.org/dr7/}}. SDSS has five bands, $u, g, r, i$, and $z$ with wavelengths 355.1, 468.6, 616.6, 748.0, and 893.2 nm, respectively. As a representative of  the optical emission we used the $g$-band.

{\bf X-ray data} (X-ray flux density, $F_{\text{X-ray}}$) are from ROSAT{\footnote{http://heasarc.gsfc.nasa.gov/docs/rosat/rosgof.html}} all-sky survey bright source catalogue  (RASS{\footnote{http://heasarc.gsfc.nasa.gov/docs/rosat/rass.html}}) and the WGA Catalog of ROSAT Point Sources (WGACAT2{\footnote{http://heasarc.gsfc.nasa.gov/wgacat/}}) at soft X-rays (0.1-2.4 keV). For sources that had both RASS and WGACAT observations we chose the closest one to the source coordinates.

{\bf Spectral data} were obtained from \citet{2006zhou1}. They are from SDSS DR3 and include: 1) the full width at half maximum of the broad H$\beta$ emission line, $FWHM$(H$\beta$), 2) the flux of the [OIII] $\lambda$5007 emission line, $F$([OIII]), 3) optical Fe II strength relative to the broad component of H$\beta$, R4570, and 4) the monochromatic luminosity at 5100\AA{}, $\lambda L_{5100}$.

The broadness of the H$\beta$ emission line can be used as a proxy for the black hole mass because the BLR clouds it arises from are gravitationally bound to the black hole. In the case of a disk-like BLR, the $FWHM$(H$\beta$) can additionally give us information about the orientation of the source. The intensity of the forbidden [O III] line arising from the NLR correlates with the underlying photoionizing continuum and can be used to estimate the intrinsic AGN power \citep[e.g.][]{1998simpson1}. The relative strength of the Fe II to the broad H$\beta$ is known to anticorrelate with the [O III] strength. This is hypothesized to be driven by the changing Eddington ratio. Increasing Eddington ratio could lead to changes in the structure of the accretion flow and thus affect the ionizing continuum seen by the NLR \citep[e.g.][]{2014shen1}. Furthermore, R4570 is possibly related to the radio emission; \citet{2008yuan1} found that the Fe II emission is on average stronger in radio-loud NLS1 galaxies than in the NLS1 population in general. $\lambda L_{5100}$ is used as an estimate of the continuum level. If a relativistic jet is present, it contributes to $\lambda L_{5100}$. However, it is not possible to estimate the level of the jet contamination with the available data.

\subsubsection{Radio-loudness parameter}
\label{sec:radioloudness}

We computed the radio-loudness ($R$) parameter for all of our sources with radio and optical data, which in total is 237 sources. We used the commonly defined $R$ value; the ratio of radio flux density ($F_{\text{R}}$) and optical flux density ($F_{\text{O}}$). For $F_{\text{R}}$ we used the radio flux density from the FIRST/NVSS survey  at 1.4~GHz and for $F_{\text{O}}$ we used the SDSS $g$-band (468.6~nm) optical flux density because it is closest to the commonly used wavelength of 440~nm. The flux density values are not simultaneous. We did not apply K-correction since it is inaccurate in this case. First, it depends on the spectral shape of the source. The spectral shape of NLS1 galaxies, especially in the radio band, varies from steep to inverted/convex in different sources, and for individual sources at different epochs \citep[e.g.][]{2017lahteenmaki1}, thus the correct K-correction is source- and time-dependent. Second, NLS1 galaxies are variable in both radio and optical bands, and this variability affects the $R$ value much more than the K-correction does; the effect of the K-correction on the radio-loudness at our $z_{max}$=0.62 is $<$30\% \citep[K-correction computed as in ][]{2011foschini1}, while the variance in $R$ induced by the variable flux density can be as high as 750\% (e.g. Gab\'{a}nyi et al. submitted). The boundary between the radio-loud and radio-quiet sources is arbitrary, and due to variability, the sources may move about the boundary, from radio-loud to radio-quiet, and vice versa. 

Even though the parameter is not necessarily the best or even an entirely correct way of estimating the radio characteristics of AGN, we use it for dividing our samples into subsamples and for comparing our results with earlier studies. According to the traditional definition 113 of our sources are radio-loud (10$ < R < $100; RL) and 27 of them are very radio-loud ($R > $100; VRL), leaving 97 radio-quiet ($R < $10; RQ) sources. Altogether, 237 of 1341 sources are radio-detected (RD). In this sample we have 1104 sources that have never been detected at radio frequencies and are therefore seemingly radio-silent (RS). However, recent radio observations performed at Mets\"{a}hovi Radio Observatory at 22 and 37~GHz indicate that at least some RS NLS1 galaxies are detectable (L\"{a}hteenm\"{a}ki et al. in prep.). Some RS NLS1 sources are clearly misclassified, but without extensive radio surveys it is impossible to say which fraction. It should be noted that the obscure boundary between RL and RQ sources, their variability, and the misclassification of some RS sources, affect the data analysis performed using these subsamples.

\subsubsection{Black hole masses}
\label{sec:bhmass}

We estimated black hole masses ($M_{\text{BH}}$) using the $FWHM$(H$\beta$) -- luminosity mass scaling relation \citep{2005greene1}

\begin{equation} M_{\text{BH}} = (4.4 \pm 0.2) \times 10^6 \bigg( \frac{\lambda L_{5100}}{10^{44} \text{ ergs s}^{-1}} \bigg)^{0.64 \pm 0.02} \bigg( \frac{\text{$FWHM$}(H\beta)}{10^3 \text{ km s}^{-1}} \bigg)^2 M_{\sun}
\label{eq:mbh} .\end{equation}

The $\lambda L_{\text{5100}}$ and $FWHM$(H$\beta$) values were taken from \citet{2006zhou1}.

We used this method instead of methods based on $M_{\text{BH}}$ -- $\sigma_{\ast}$ (stellar velocity dispersion of the bulge) or the $M_{\text{BH}}$ -- $M_{\text{bulge}}$ (the mass of the bulge) relation \citep{2009bentz1} because $\sigma_{\ast}$ or $L_{\text{bulge}}$ data are not available for most NLS1 sources. They also tend to lie below the normal $M_{\text{BH}} - \sigma_{\ast}$ and $M_{\text{BH}} - L_{\text{bulge}}$ relations \citep{2001mathur1, 2001laor1, 2006zhou1}. However, this method does not take into account possible inclination effects caused by the geometry of the BLR and the viewing angle \citep{2011decarli2}, and for sources with relativistic jets this method overestimates the black hole mass \citep{2004wu1} because the jet contamination to $\lambda L_{\text{5100}}$ cannot be estimated due to lack of data. The method is not precise but can be used as an order of magnitude estimate in statistical studies. We were able to estimate $M_{\text{BH}}$ for 1312 sources.

\subsection{Environment data samples}
\label{sec:envdata}

We use three sets of environment data in our analyses. 

{\bf SDSS Main Galaxy sample and luminosity-density field} (LDF). In this study we use the catalogue of NLS1 galaxies by \citet{2006zhou1}, constructed from the spectroscopic sample of the SDSS Data Release 3 and consisting of 2011 NLS1 sources with redshifts $z \leq 0.8$. We cross-matched these galaxies with the galaxy and group catalogues by \citet{2014A&A...566A...1T} and filament catalogues by \citep{2014tempel1} and found 229 matches with redshifts $z \leq 0.2$. There remain $\sim$300 NLS1 sources with $z < 0.2$ that could not be matched. The galaxy catalogue by \citet{2014A&A...566A...1T} is based on the SDSS Data Release 10 \citep{2014ahn1} and uses only the main contiguous area of the SDSS Legacy Survey. It consists of 588193 galaxies and 82458 groups spanning an area of 7221 square degrees in the sky with redshifts $z \leq 0.2$. It also provides luminosity-density estimates with 1, 2, 4 and 8 $h^{-1}$ Mpc smoothing for each galaxy and can be used to study the effect of environment on NLS1 galaxies. The redshift range of this sample is $z$ = 0.0133---0.1987 and the mean $z = 0.1078$.

{\bf SDSS LRG LDF} is a three-dimensional low-resolution LDF constructed using a sample of luminous red galaxies (LRG) in SDSS Data Release 7 \citep{2009abazajian1}. The mean density around each source is calculated in a volume of 3 $h^{-1}$ Mpc, giving an estimate of its surroundings on a supercluster-scale. 16 $h^{-1}$ Mpc smoothing was used. This field is limited to 225--1000 $h^{-1}$ Mpc. For a detailed description, see Appendix~\ref{densityfield}, \citet{2011lietzen1}, and \citet{2012liivamagi1}. We have SDSS LRG LDF data for 960 sources. The mean redshift for our sample is $z = 0.2340$, minimum $z = 0.0726$, and maximum $z = 0.3996$. This sample and the SDSS Main Galaxy sample have 171 overlapping NLS1 sources.

{\bf SDSS BOSS LDF} is constructed using the SDSS Baryon Oscillation Spectroscopic Survey (BOSS) Constant MASS (CMASS) sample from Data Release 12 \citep{2015alam1}, limited to 1200--1600 $h^{-1}$ Mpc ($z=0.43-0.62$). The field was calculated in a 3 $h^{-1}$ Mpc grid, with a 8 $h^{-1}$ Mpc smoothing. For a detailed description see Appendix~\ref{densityfield} and  \citet{2016lietzen1}. We have these data for 323 sources. The mean, minimum, and maximum redshifts for our sample are $z = 0.5224, z = 0.4412$ and $z = 0.6193$, respectively.

The group and filament data are available only for the Main Galaxy sample. The LDF, or for short, density, data is available for all samples but are not straightforwardly comparable with each other because of different smoothing scales and different selection criteria for the galaxies. The LDF is based on mean densities, not the actual physical large-scale structures. Because of this, the regions with intermediate densities can be either real physical filaments or transition zones between superclusters and voids. From now on the whole sample refers to all sources for which we have any environmental data, that is a total of 1341 sources.

\begin{figure*}[ht!]
\centering
\includegraphics[width=0.99\textwidth]{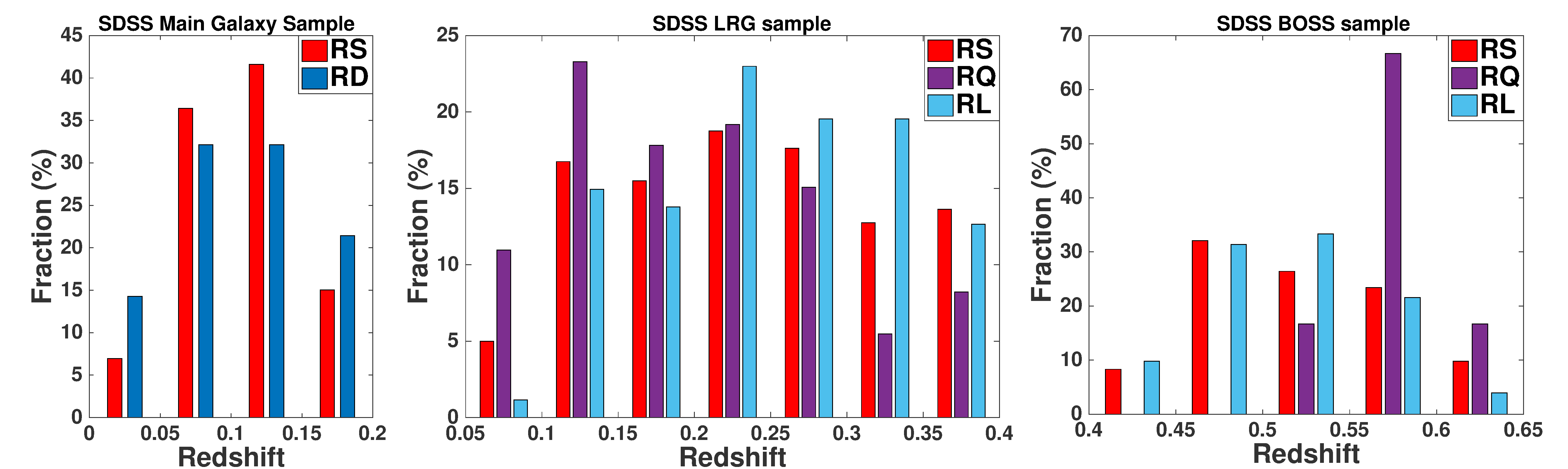}
\caption{Redshift distributions of the three samples. \emph{Left.} SDSS Main Galaxy sample, 229 sources. \emph{Middle.} SDSS LRG sample, 960 sources, of which 171 overlap with the SDSS Main Galaxy
sample. \emph{Right.} SDSS BOSS sample, 323 sources.}
\label{fig:zdist}
\end{figure*}

The redshift distributions of the samples are shown in Figure~\ref{fig:zdist} and the average redshifts are given in Table~\ref{tab:zmbh}. RL sources lie on average at higher redshifts than RQ sources. This might be due to either an observational bias or evolution of AGN. It is unlikely that this could be caused by an optical observational bias since DR3, based on which most of our sources were selected, is 95\% complete down to $m_g$ = 22.2{\footnote{http://classic.sdss.org/dr3/}}, and most NLS1 galaxies are much brighter than this. Most of our radio data are from the FIRST survey which is complete down to 1~mJy, so we might miss a population of faint non-detectable radio emitters incorrectly classified as RS, but it is improbable that this could explain the whole difference. With future observatories we will be able to detect an abundant population of very radio-faint NLS1 galaxies. However, in general, these sources are expected to be optically faint and difficult to classify as NLS1 galaxies \citep{2015berton2}, and they probably do not contaminate our sample, either. Thus the evolution of AGN is a more plausible explanation; AGN activity was more prominent in the earlier Universe \citep[e.g.][]{2005wolf1,2012beckmann1}. Therefore it is consistent that RL sources have higher redshifts.

\begin{figure}[ht!]
\centering
\includegraphics[width=0.49\textwidth]{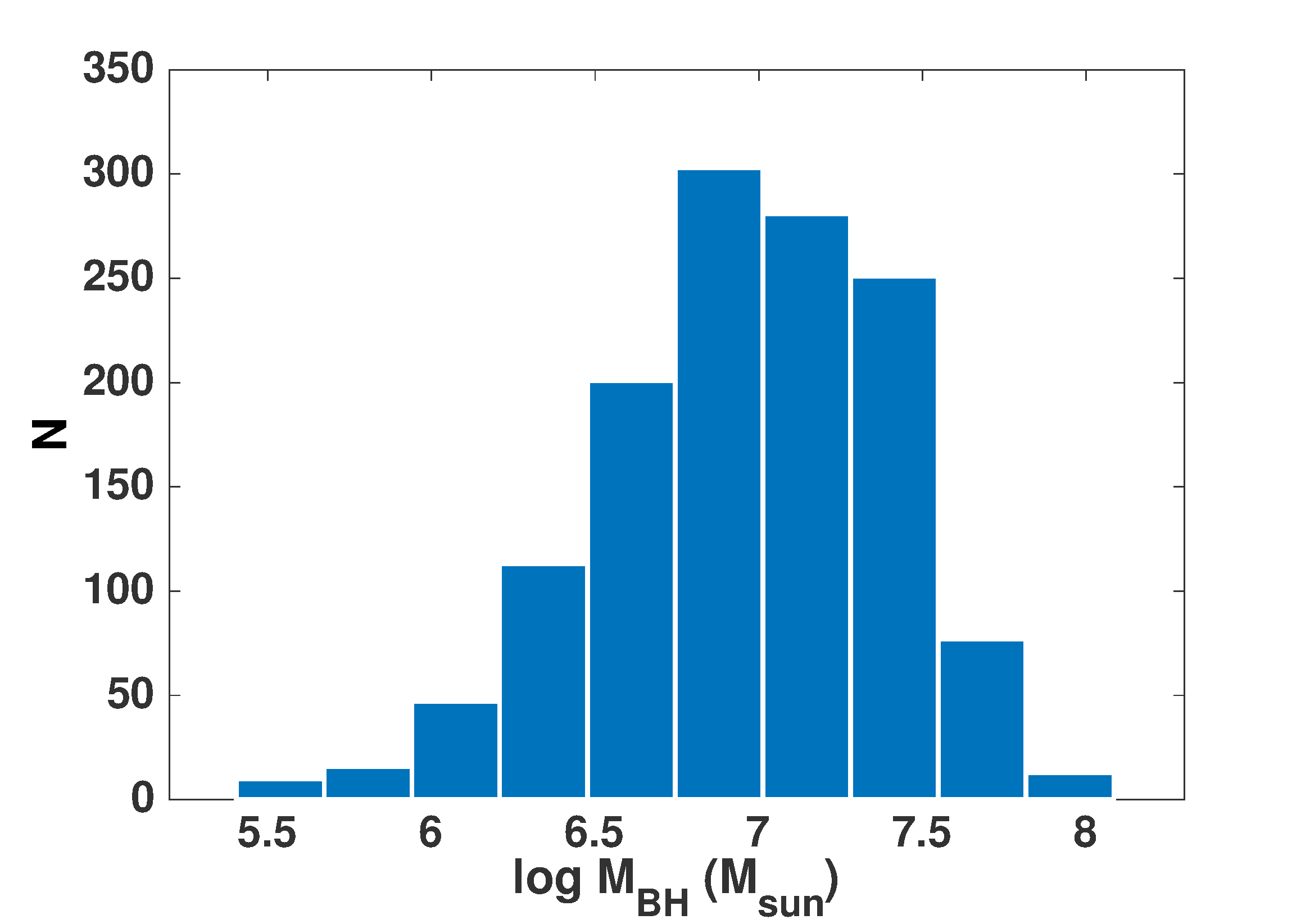}
\caption{Black hole mass distribution of the whole sample of 1341 sources.}
\label{fig:mbhdist}
\end{figure}

The average $M_{\text{BH}}$ values for the samples are given in Table~\ref{tab:zmbh} and the black hole mass distribution of the whole sample is shown in Figure~\ref{fig:mbhdist}. At the closest redshift range in the SDSS Main Galaxy Sample the RD sources harbour slightly more massive black holes than RS sources. In the SDSS LRG sample there is a small difference between the RS and RD sources, but no difference between RQ and RL sources. At the farthest redshift range, RS sources have the lowest black hole masses, while RL sources have masses only a little higher. RQ sources have the highest average black hole mass, but the sample size is small. The trend of increasing black hole mass with higher redshift is clear and is caused by the correlation of $\lambda L_{5100}$ and redshift, that is, NLS1 galaxies are brighter farther away.

\begin{table*}[ht]
\caption[]{Average redshifts and black hole masses of different subsamples.}
\centering
\begin{tabular}{p{1.6cm} | p{0.35cm} p{1.95cm} l | p{0.35cm} p{1.95cm} l | p{0.35cm} p{1.95cm} l}
\hline\hline
           &  \multicolumn{3}{c|}{SDSS MGS}                             &   \multicolumn{3}{c|}{SDSS LRG}                            &   \multicolumn{3}{c}{SDSS BOSS}                   \\
Sample     & $N$   & $z$               & log $M_{\text{BH}}$ ($M_{\sun}$) & $N$   & $z$               & log $M_{\text{BH}}$ ($M_{\sun}$) & $N$   & $z$       & log $M_{\text{BH}}$ ($M_{\sun}$)  \\ \hline
all        & 229 & 0.108 $\pm$ 0.003 & 6.57 $\pm$ 0.03                  & 960 & 0.234 $\pm$ 0.003 & 6.86 $\pm$ 0.01                  & 323 & 0.522 $\pm$ 0.003 & 7.37 $\pm$ 0.02   \\
RS         & 173 & 0.109 $\pm$ 0.003 & 6.55 $\pm$ 0.03                  & 799 & 0.235 $\pm$ 0.003 & 6.85 $\pm$ 0.01                  & 266 & 0.523 $\pm$ 0.003 & 7.35 $\pm$ 0.02   \\
RD         & 56  & 0.103 $\pm$ 0.006 & 6.63 $\pm$ 0.05                  & 161 & 0.227 $\pm$ 0.007 & 6.91 $\pm$ 0.03                  & 57  & 0.522 $\pm$ 0.007 & 7.47 $\pm$ 0.04   \\
RQ         &     &                   &                                  & 73  & 0.201 $\pm$ 0.010 & 6.91 $\pm$ 0.05                  & 6   & 0.577 $\pm$ 0.010 & 7.72 $\pm$ 0.09   \\
RL         &     &                   &                                  & 74  & 0.249 $\pm$ 0.009 & 6.91 $\pm$ 0.04                  & 37  & 0.514 $\pm$ 0.009 & 7.46 $\pm$ 0.05   \\
VRL        &     &                   &                                  & 13  & 0.247 $\pm$ 0.026 & 6.96 $\pm$ 0.10                  & 14  & 0.521 $\pm$ 0.013 & 7.36 $\pm$ 0.07   \\
RL + VRL   &     &                   &                                  & 87  & 0.249 $\pm$ 0.009 & 6.91 $\pm$ 0.04                  & 51  & 0.515 $\pm$ 0.007 & 7.44 $\pm$ 0.04    \\
XD         & 149 & 0.105 $\pm$ 0.004 & 6.58 $\pm$ 0.03                  & 418 & 0.219 $\pm$ 0.004 & 6.87 $\pm$ 0.02                  & 97  & 0.522 $\pm$ 0.006 & 7.39 $\pm$ 0.03 \\ \hline

\end{tabular}
\label{tab:zmbh}
\end{table*}

\section{Results}
\label{sec:results}

In this section we present the results of the various statistical studies we performed to investigate the impact of the large-scale environment. The large-scale environment analysis of NLS1 galaxies was done separately for the three large-scale environment density data samples and redshift ranges, and is presented in ~\ref{sec:maingalaxy},~\ref{sec:sdsslrg}, and ~\ref{sec:sdssboss}. We studied the changes in the large-scale environments of subsamples of NLS1 galaxies, and compare them with other AGN classes. We performed principal component analysis (PCA) for the whole sample and selected subsamples. PCA was run separately for the SDSS LRG (~\ref{sec:lrgpca}) and BOSS (~\ref{sec:bosspca}) samples. In Section \ref{sec:evcorr} we study how the principal components are correlated with the intrinsic properties of NLS1 sources.

\subsection{Large-scale environment}
\label{sec:density}

\subsubsection{SDSS Main Galaxy sample}
\label{sec:maingalaxy}

This sample with 229 sources is rather small so we divided it into only two subsamples, RS and RD, to obtain more reliable statistics. The averages of the luminosity-density with the 8 $h^{-1}$ Mpc smoothing scale, the group richness, the group radii and the distance to the closest filament axis for the whole sample and the subsamples are shown in Table~\ref{tab:sdssmain}. The error is the standard error of the mean. The data are not complete; the number of sources used to calculate each average value is given in Table~\ref{tab:sdssmain} in the `$N$' column. The average luminosity-densities for the subsamples, using any of the available smoothing scales, are similar, as is the average richness of the groups NLS1 galaxies reside in. Although only 52 out of 229 sources belong to a group ($N_{\text{gal}} \geq$ 3) , 22 are in pairs and 155 are field galaxies. The average black hole mass in isolated NLS1 galaxies is slightly higher (6.63$\pm$0.03) than in NLS1 galaxies in pairs or in groups (6.45$\pm$0.05). It should be noted that most of the NLS1 sources reside in poor groups, and a few individuals in unusually rich clusters dominate the average. Two of our sources reside in the same cluster, so average $N_{\text{gal}}$, $R_{\text{vir}}$ and $R_{\text{max}}$ are calculated using 51 groups. The average radii, $R_{\text{vir}}$ and $R_{\text{max}}$, of the groups are typical for groups of this size \citep{1996bahcall1}. RS and RD sources are similar in their properties except for the size of the groups they are in; RS NLS1 galaxies reside in groups with larger virial/maximum radii than RD sources. Since the average richness of the groups is similar, this means that the groups that RD sources reside in have a higher number density. In the SDSS filament catalogue we use \citep{2014tempel1}, the filament radius is fixed to $R_{\text{F}}$ = 0.5 h$^{-1}$Mpc. Out of 157 NLS1 galaxies with filament data, 40 have $D_{\text{F}}$ < 0.5 h$^{-1}$Mpc, thus most of the sources in our sample reside in a void.

\begin{table*}[ht]
\caption[]{SDSS Main Galaxy sample NLS1 statistics}
\centering
\begin{tabular}{l | l l | l l l l | l l}
\hline\hline
& $N$ & Average density   & $N$  & $N_{\text{gal}}$\tablefootmark{a} & $R_{\text{vir}}$\tablefootmark{b} (h$^{-1}$Mpc) & $R_{\text{max}}$\tablefootmark{c} (h$^{-1}$Mpc) & $N$ & $D_{\text{F}}$ (Mpc)\tablefootmark{d} \\ \hline
all & 229 & 2.60 $\pm$ 0.13   & 51 & 10.96 $\pm$ 1.92           & 0.28 $\pm$ 0.02                          & 0.50 $\pm$ 0.05                          & 157 & 2.74 $\pm$ 0.20   \\
RS  & 173 & 2.61 $\pm$ 0.15   & 41 & 10.44 $\pm$ 1.99           & 0.29 $\pm$ 0.02                          & 0.53 $\pm$ 0.06                          & 119 & 2.75 $\pm$ 0.23   \\
RD  & 56  & 2.56 $\pm$ 0.27   & 10 & 12.91 $\pm$ 5.45           & 0.23 $\pm$ 0.05                          & 0.41 $\pm$ 0.13                          &  38 & 2.71 $\pm$ 0.40    \\ \hline
\end{tabular}
\label{tab:sdssmain}
\tablefoot{
\tablefoottext{a}{Average richness of a group.}
\tablefoottext{b}{Virial radius of the group, projected harmonic mean.}
\tablefoottext{c}{Maximum radius of the group, projected harmonic mean.}
\tablefoottext{d}{Distance to the closest filament axis.}
}
\end{table*}

\subsubsection{SDSS LRG LDF}
\label{sec:sdsslrg}

\begin{figure}[ht!]
\centering
\includegraphics[width=0.51\textwidth]{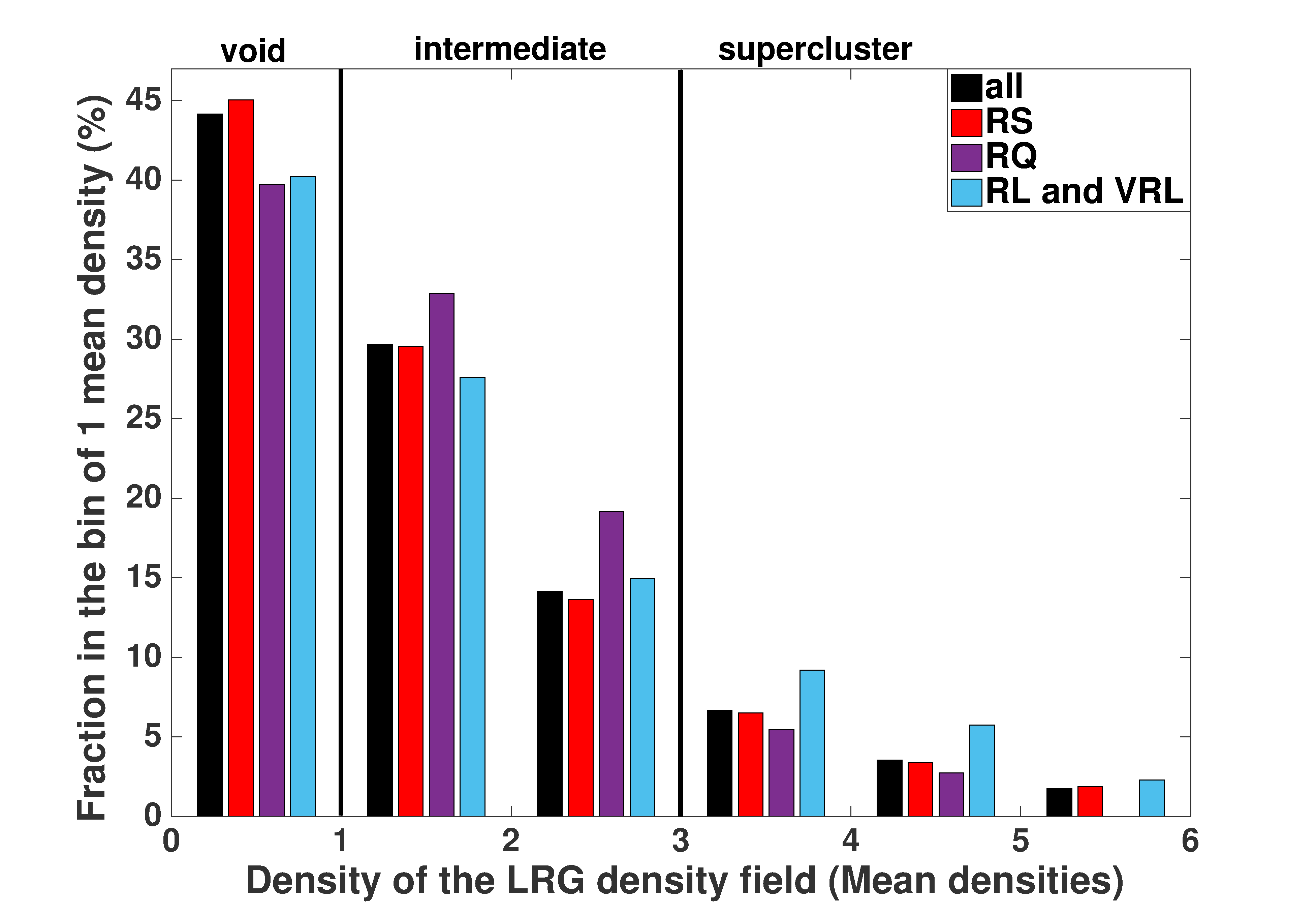}
\caption{Distributions of the whole sample and RQ, RQ, and RL plus VRL subsamples of NLS1 galaxies in the different environments defined with the 
SDSS LRG LDF. The last bin (5--6) includes all sources with density $>$5.}
\label{fig:densityhist}
\end{figure}

We have SDSS LRG LDF data for 960 sources, of which 171 are included in the SDSS Main Galaxy sample as well. We are interested in how the subsamples and NLS1 galaxies in general are distributed in the large-scale enviroment, and if the distribution is similar to the distributions of Seyfert galaxies and other types of AGN in \citet{2011lietzen1}. Density distributions for the whole sample, RS, RQ, and RL plus VRL subsamples are shown in Figure~\ref{fig:densityhist}. Table~\ref{tab:denspros} shows average densities for the subsamples and the fraction of sources in voids, intermediate density regions, and superclusters. The error is the standard error of the mean. Sy1 and Sy2 galaxies from \citet{2011lietzen1} are listed for comparison. We use the same definition for voids, intermediate density regions, and superclusters as in \citet{2011lietzen1}; in a void the luminosity-density is less than 1.0, in an intermediate-density region the luminosity-density is between 1.0 and 3.0, and in a supercluster the luminosity-density is larger than 3.0. These limits were defined by spatially overlapping LRG LDF with the SDSS Main Sample LDF \citep{2011lietzen1}. As mentioned before, the intermediate luminosity-density regions defined this way might correspond to either physical filaments or other areas of intermediate mean density; for example, the boundaries between superclusters and voids.

The average density for the whole sample is smaller than for any sample in \citet{2011lietzen1}, and the distribution to voids, intermediate-density regions, and superclusters is clearly
distinct compared to the samples in \citet{2011lietzen1}. These results imply that the large-scale environment of NLS1 galaxies is different when compared to other AGN.
The difference from Sy1 and Sy2 galaxies in the average densities is not so pronounced, but is significant nonetheless. This divergence is supported by the differences in their spatial distributions; there are more NLS1 galaxies than Sy1 and Sy2 galaxies in voids, and less in intermediate-density regions. Interestingly the fraction is almost the same in superclusters.

Amongst the NLS1 subsamples the average density increases with increasing radio loudness, and the distribution of sources to voids, intermediate-density regions, and superclusters changes with 
varying radio loudness. For RL and RQ sources the fraction in voids is the same (40\%), but a larger percentage of RQ than RL sources reside in intermediate-density regions (52\% vs. 43\%), and 
vice versa for superclusters (8\% vs. 16\%). The fraction of the RS sources residing in voids (45\%) and intermediate-density regions (43\%) is very high. 
But surprisingly, a bigger fraction of RS than RQ sources reside in superclusters (12\% vs. 8\%).
This might indicate, and be explained, by the misclassification of a fraction of RS sources, meaning that the RS subsample is mixed. The two-sample Kolmogorov-Smirnov (KS) test also suggests this; 
according to the KS-test the density distributions of our subsamples -- RS vs. RD, RQ vs. RL and RS vs. RL -- are drawn from the same distribution.

\citet{2011lietzen1} studied the large-scale environments of various types of AGN using the same SDSS LRG LDF that is used in this paper. Their sample included subsamples of radio-quiet (radio luminosity, $L_{\text{R}} <$ 10$^{25}$ W Hz$^{-1}$) and radio-loud ($L_{\text{R}} >$ 10$^{25}$ W Hz$^{-1}$) quasars, BL Lac objects, Sy1 and Sy2 galaxies, and radio galaxies divided into flat-spectrum, FR I, and FR II sources. Their Seyfert samples do not include NLS1 galaxies since they only chose sources with [OIII]/H$\beta >$ 3, and for NLS1 sources [OIII]/H$\beta <$ 3 by definition. Their main result was that radio galaxies tend to reside in denser large-scale environments than radio-quiet quasars, which can be explained by the AGN evolution scenario.

\begin{table*}[ht]
\caption[]{Average density of the whole SDSS LRG sample and the NLS1 subsamples, and percentage in voids, intermediate density regions, and superclusters.}
\centering
\begin{tabular}{l l l l l l}
\hline\hline
                              & N     & Average density & LD\tablefootmark{a}<1 (\%)  & 1<LD<3 (\%) & LD>3 (\%) \\ \hline
all                           & 960   & 1.50 $\pm$ 0.04 & 44        & 44         & 12                \\
RS                            & 799   & 1.48 $\pm$ 0.05 & 45        & 43         & 12                \\
RD                            & 161   & 1.61 $\pm$ 0.10 & 40        & 47         & 13                \\
RQ                            & 73    & 1.48 $\pm$ 0.08 & 40        & 52         & 8                 \\
RL                            & 74    & 1.66 $\pm$ 0.16 & 41        & 43         & 16                \\
VRL                           & 13    & 2.01 $\pm$ 0.36 & 38        & 38         & 23                \\ 
RL and VRL                    & 87    & 1.71 $\pm$ 0.14 & 40        & 43         & 17                \\ \hline
Sy1 galaxies\tablefootmark{b} & 1095  & 1.73 $\pm$ 0.04 & 34        & 51         & 15                \\
Sy2 galaxies\tablefootmark{b} & 2494  & 1.65 $\pm$ 0.03 & 35        & 52         & 13                \\ \hline

\end{tabular}
\label{tab:denspros}
\tablefoot{
\tablefoottext{a}{LD = mean luminosity-density of the LDF.}
\tablefoottext{b}{\citet{2011lietzen1}}
}
\end{table*}

{\bf Overlap of Main Galaxy sample and LRG LDF.} There are 171 overlapping sources in the SDSS Main Galaxy sample and LRG LDF. With this subset of sources we can study how the physical properties derived from the SDSS Main Galaxy sample correlate with the SDSS LRG LDF. The average densities of the LRG LDF for the whole overlapping subset and subsamples defined by the group richness are shown in Table~\ref{tab:overlap}, and the correlation results between the Main Galaxy sample properties and LRG luminosity-density are presented in Table~\ref{tab:main-lrg-corr}.

Table~\ref{tab:overlap} and the significant correlation in Table~\ref{tab:main-lrg-corr} clearly show that groups with more members reside in denser large-scale regions. A similar trend was found in \citet{2017poudel1}. The LRG luminosity-density (16 h$^{-1}$Mpc smoothing scale) correlates with the Main Galaxy sample luminosity-density with the smoothing scale of 8 h$^{-1}$Mpc, but not with the 1 h$^{-1}$Mpc smoothing scale, which is of course due to the differences in the smoothing scales. In addition, the LRG luminosity-density correlates with the total group mass, which is due to the increasing group richness towards the denser large-scale environment.

\begin{table}[ht]
\caption[]{Average LRG luminosity-densities of the overlapping sources of the SDSS Main Galaxy sample and LRG, and subsamples defined by the group richness.}
\centering
\begin{tabular}{l l l}
\hline\hline
                                       & N   & Average density \\ \hline
all                                    & 171 & 1.57 $\pm$ 0.08 \\
$N_{\text{gal}}$\tablefootmark{a} = 1  & 122 & 1.44 $\pm$ 0.09 \\
$N_{\text{gal}}$ > 1                   & 49  & 1.88 $\pm$ 0.15 \\
$N_{\text{gal}}$ > 2                   & 34  & 1.99 $\pm$ 0.19 \\
$N_{\text{gal}}$ > 10                  & 10  & 2.72 $\pm$ 0.31 \\ \hline

\end{tabular}
\label{tab:overlap}
\tablefoot{
\tablefoottext{a}{Group richness.}
}
\end{table}

\begin{table}[ht]
\caption[]{Spearman rank correlations and probability values (in parentheses) for the overlapping sources of the SDSS Main Galaxy sample and SDSS LRG sample. Correlations in boldface have p$<$0.05.}
\centering
\begin{tabular}{l l}
\hline\hline
                                   & LRG LD  \\ \hline
LD 1\tablefootmark{a}              &  0.049 (0.529)   \\
LD 8\tablefootmark{b}              &  {\bf 0.765} (0.000)   \\
$N_{\text{gal}}$\tablefootmark{c}  &  {\bf 0.226} (0.003)   \\
$D_{\text{F}}$\tablefootmark{d}    & -0.003 (0.974)   \\
$R_{\text{vir}}$\tablefootmark{e}  &  0.150 (0.309)   \\
$M_{NFW}$\tablefootmark{f}         &  {\bf 0.288} (0.048)   \\ \hline

\end{tabular}
\label{tab:main-lrg-corr}
\tablefoot{
\tablefoottext{a}{Luminosity-density, 1 h$^{-1}$Mpc smoothing scale.}
\tablefoottext{b}{Luminosity-density, 8 h$^{-1}$Mpc smoothing scale.}
\tablefoottext{c}{Group richness.}
\tablefoottext{d}{Distance to the closest filament axis, Mpc.}
\tablefoottext{e}{Virial radius of the group, projected harmonic mean, h$^{-1}$Mpc}
\tablefoottext{f}{Total mass of the group, NFW profile, h$^{-1}$Mpc}
}
\end{table}

\subsubsection{SDSS BOSS CMASS LDF}
\label{sec:sdssboss}

We have SDSS BOSS CMASS LDF data for 323 sources. There are no overlapping sources with the SDSS Main Galaxy sample or the SDSS LRG LDF since the redshift range is different (0.43---0.62). The results for the BOSS LDF sample are displayed in Table~\ref{tab:bossdenspros}. The average densities of the RQ and RL subsamples are significantly different, even when taking into account the comparatively large errors{\footnote{Standard error of the mean.}}. None of the RQ sources reside in a supercluster, although the RQ subsample contains only six sources and is therefore probably not a good representative of the overall RQ population. 16\% of RL sources are located in superclusters. Interestingly the average density of RS sources lies between those of RQ and RL sources, and 7\% of RS sources reside in supercluster environments. This apparent contradiction might result from the small sample size of RQ sources and the mixed nature of the RS subsample. However, in both LRG and BOSS LDFs, the fraction of RS sources in superclusters is larger than the fraction of RQ sources. The main result of the BOSS LDF analysis agrees with the LRG LDF analysis; the radio-loudness increases with the enviromental density.

Due to different smoothing scales, the mean density values of the LRG and BOSS LDFs are not straightforwardly comparable, and there are no previous studies of the average densities of other types of AGN in the BOSS LDF. In the BOSS LDF only the boundary between a supercluster and the outside of a supercluster has been defined. This was done by matching the supercluster volumes with the SDSS Main Sample superclusters \citep{2016lietzen1}. It is noteworthy that at this redshift range there is a clear deficit of RQ sources and an excess of RL sources when compared to the LRG sample. This issue was discussed in Sect.~\ref{sec:envdata}.

We also compared the average densities of NLS1 galaxies with those of their candidate parent population samples from \citet{2015berton1}. Unfortunately we were only able to obtain the large-scale environment data for six sources, rendering the samples statistically insignificant. 

\begin{table*}[ht]
\caption[]{Average density of the whole SDSS BOSS sample and the NLS1 subsamples, and the fraction outside of and in superclusters.}
\centering
\begin{tabular}{l l l l l l}
\hline\hline
            & N     & Average density & LD\tablefootmark{a}<6 (\%)  & LD>6 (\%)\\ \hline
all         & 323   & 2.29 $\pm$ 0.15 & 92        & 8       \\
RS          & 266   & 2.13 $\pm$ 0.15 & 93        & 7       \\
RD          & 57    & 3.05 $\pm$ 0.54 & 86        & 14      \\
RQ          & 6     & 1.12 $\pm$ 0.51 & 100       & 0       \\
VRL         & 14    & 2.35 $\pm$ 0.64 & 93        & 7       \\
RL and VRL  & 51    & 3.27 $\pm$ 0.60 & 84        & 16      \\  \hline

\end{tabular}
\label{tab:bossdenspros}
\tablefoot{
\tablefoottext{a}{LD = mean luminosity-density of the LDF.}
}
\end{table*}

\subsection{PCA}
\label{sec:pca}

PCA is a statistical method used to simplify large amounts of data, and to study the underlying correlations that do not show in basic correlation analyses, and might include multiple parameters. Using an orthogonal transformation it converts a set of possibly correlated variables into a set of linearly uncorrelated variables called principal components, or eigenvectors (EV). The first principal component accounts for as much of the variability in the data as possible. The second principal component has as large a variance as possible while still being orthogonal to the first principal component, and so on. This method makes it possible to find underlying connections and the most dominant variables in a data set, and possibly helps to identify the physical properties connected with each EV. While PCA is a powerful tool to study and systematise extensive data sets it should be noted that the results of PCA always depend on the set of parameters and the sample used. A good explanation and overview of the PCA can be found in \citet{2010abdi1}.

\citet{1992boroson1} used PCA to study the optical properties of 87 quasi-stellar objects (QSO); they found the EV1 to be dominated by the anticorrelation between the strength of Fe II, and the strength of [O III] $\lambda$5007 and $FWHM$(H$\beta$), while EV2 distinguished between the strength of He II $\lambda$4686 and optical luminosity. In \citet{2002boroson1}, 75 sources were added to the original sample, yielding consistent results. They suggested that the EV1 correlations are driven by the Eddington ratio, $L/L_{\text{Edd}}$, and EV2 by the accretion rate. \citet{2012xu1} studied a sample of NLS1 and broad-line Seyfert 1 galaxies (BLS1) using PCA, and found their EV1 and EV2 to be in good agreement with \citet{1992boroson1} and \citet{2002boroson1}. \citet{2004grupe1} performed PCA for a sample of 110 soft-X-ray-selected AGN -- of which about half were NLS1 galaxies -- and found the EV1 to be similar to the EV1 in \citet{2002boroson1}, and the EV2 to strongly correlate with the black hole mass. In \citet{2015jarvela1} we used PCA to study a pure sample of 292 NLS1 sources; in our study the EV1 was dominated by the anticorrelation between $F_{\text{O}}$ and $F_{\text{IR}}$, and M$_{\text{BH}}$ and $FWHM$(H$\beta$), and it did not correlate with the Eddington ratio. In contrast, our EV2 was the `traditional' EV1, distinguishing between R4570 and $FWHM$(H$\beta$), and correlating strongly with the Eddington ratio. The correlation space defined by $FWHM$(H$\beta$), $F$([OIII]) $\lambda$5007, R4570, and C IV $\lambda$1549 has been established as the 4DE1 parameter space explaining a host of differences observed in AGN, and possibly tracing the general AGN evolution \citep{2006marziani1, 2007sulentic1}.

We performed weighted PCA separately for the LRG and the BOSS samples using the \texttt{pca}{\footnote{https://se.mathworks.com/help/stats/pca.html}} function in MATLAB Statistics and Machine Learning Toolbox. This is the first time a parameter describing the large-scale environment has been used in the PCA of AGN, allowing us to study its connection to the AGN correlation space. The variables were selected so that there exists no strong correlations between them, because including already correlated variables to PCA skews the results. The variables used in all PC analyses were $F_{\text{O}}$, $FWHM$(H$\beta$), R4570, $F$([O III]), and the large-scale environment density, either from LRG or BOSS. Values for $FWHM$(H$\beta$), R4570, and $F$([O III]) were taken from \citet{2006zhou1}. In addition to these variables we used $F_{\text{R}}$ in the PCA of RD sources and $F_{\text{X-ray}}$ in the PCA of an X-ray-detected (XD) sample. With these additional samples we wanted to study whether the correlation space is consistent among samples with disparate selection criteria; the diversity of the RD sample should be kept in mind. We excluded the infrared wavebands since they are tightly correlated to the optical emission and $M_{\text{BH}}$ which is correlated with $FWHM$(H$\beta$).

The results are shown in Tables~\ref{tab:lrgpcaev1}-\ref{tab:bosspcaev3}, and in Figures~\ref{fig:biplotlrg} and~\ref{fig:biplotboss}. In the tables, the coefficients have been grouped together based on their sign, that is, whether they correlate or anticorrelate with the EV and each other. In this way, it is easier to see which properties dominate each EV and what  their mutual relations are. We used the same notation in \citet{2015jarvela1}. This information is also presented in biplots; the direction and the length of each vector corresponds to the level of contribution of a variable to an EV, and shows the relations between the variables. The direction of the coefficients is arbitrary, that is, the sign of the coefficient is insignificant as such and only their respective direction matters.

\subsubsection{LRG PCA}
\label{sec:lrgpca}

\begin{figure*}[ht!]
\centering
\includegraphics[width=0.99\textwidth]{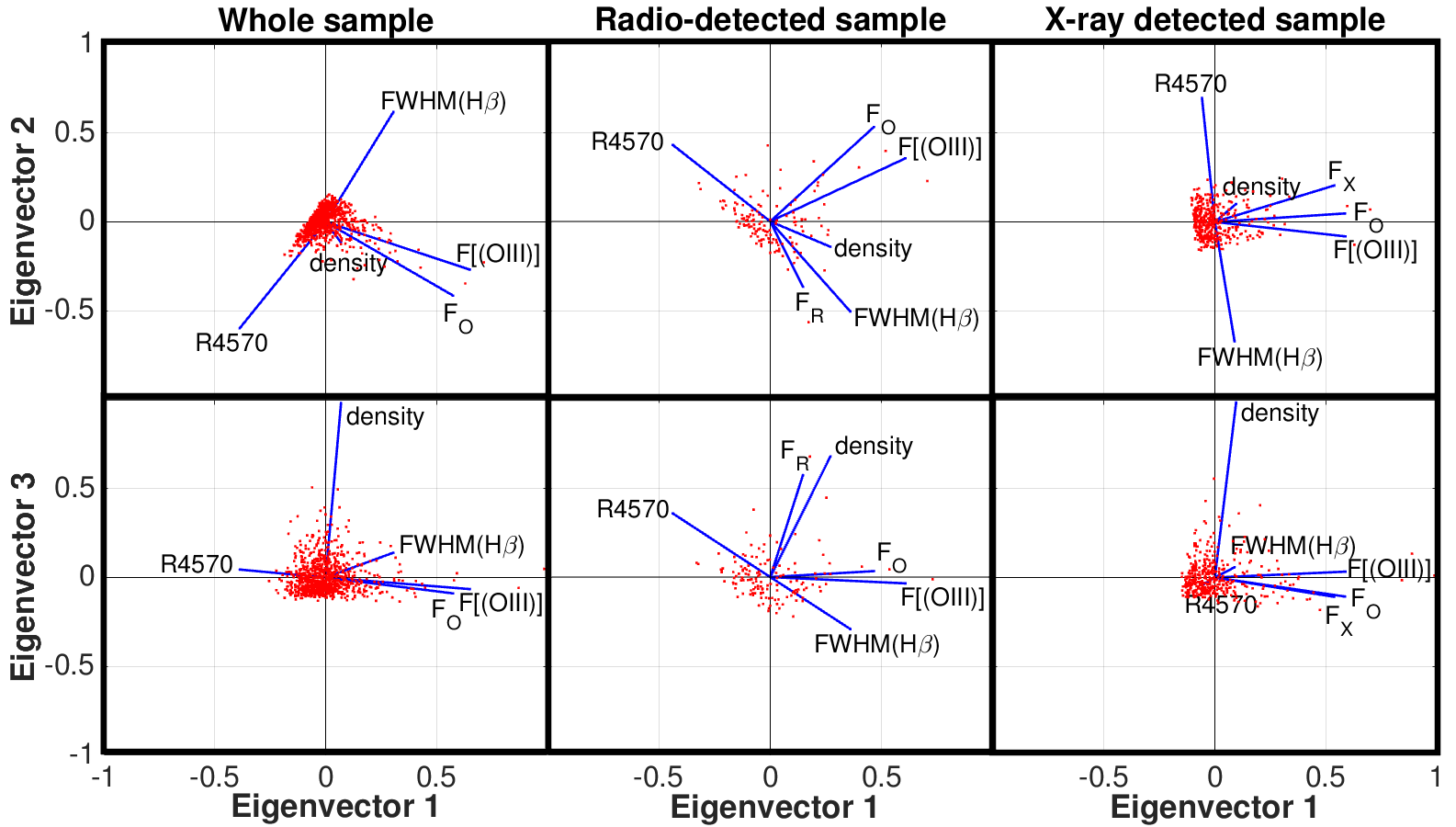}
\caption{Biplots of EVs 1, 2, and 3, and variables of the LRG PCA.}
\label{fig:biplotlrg}
\end{figure*}

We performed the LRG PCA with those sources that have all the necessary data (935 sources), and with two samples with $F_{\text{R}}$ (139 sources) or $F_{\text{X-ray}}$ (408 sources) as an additional parameter. The results for the first three principal components are displayed and the PCA coefficients listed in Tables~\ref{tab:lrgpcaev1},~\ref{tab:lrgpcaev2}, and ~\ref{tab:lrgpcaev3}, and the corresponding biplots are shown in Figure~\ref{fig:biplotlrg}.

{\bf EV1} accounts for 26\% -- 32\% of the variance. In all samples EV1 is dominated by $F_{\text{O}}$ and $F$([OIII]), which in the whole and RD samples are anticorrelated with R4570. $F_{\text{R}}$ does not play a significant role in EV1 of the RD sample. In the XD sample, R4570 is insignificant, and $F_{\text{X-ray}}$ correlates with $F_{\text{O}}$ and $F$([OIII]). The EV1 of this pure NLS1 sample is not exactly similar to the traditional EV1 distinguishing between $FWHM$(H$\beta$) and $F$([OIII]), and R4570, although $FWHM$(H$\beta$) slightly contributes to EV1 in the whole and RD samples. Interestingly, in the XD sample, $FWHM$(H$\beta$) is insignificant.

\begin{table}[ht]
\begin{minipage}{0.5\textwidth}
\caption{Results of the LRG PCA: EV1. The coefficients have been grouped together based on their sign.}
\centering
\begin{tabular}{l l l}

\hline\hline
-            & Sample     & + \\ \hline
R4570 -0.39  & All (30\%) & $F$([OIII]) 0.65         \\ 
             &            & $F_{\text{O}}$ 0.57           \\
             &            & $FWHM$(H$\beta$) 0.31 \\
             &            & density 0.07  \\
 & & \\
R4570 -0.44  & RD (26\%)  & $F$([OIII]) 0.61 \\
            &                        & $F_{\text{O}}$ 0.47 \\
            &                        & $FWHM$(H$\beta$) 0.36 \\
            &                        & density 0.27 \\
            &                        & $F_{\text{R}}$ 0.17 \\
& & \\
R4570 -0.06    & XD (32\%)      & $F$([OIII]) 0.59 \\
               &                       & $F_{\text{O}}$ 0.59 \\ 
               &                       & $F_{\text{X-ray}}$ 0.54 \\
               &                       & density 0.10 \\ 
               &                       & $FWHM$(H$\beta$) 0.09 \\ \hline

\end{tabular}
\label{tab:lrgpcaev1}
\end{minipage}
\end{table}

{\bf EV2} 23\% -- 25\% of the variance is explained by EV2, which is clearly dominated by the anticorrelation of $FWHM$(H$\beta$) and R4570. In the whole and RD samples $F_{\text{O}}$ correlates with R4570, but in the XD sample it is negligible. Since EV2 is led by the R4570 -- $FWHM$(H$\beta$) anticorrelation, it is probably similar to EV1 found in \citet{1992boroson1, 2002boroson1} and \citet{2012xu1}.

\begin{table}[ht]
\begin{minipage}{0.5\textwidth}
\caption{Results of the LRG PCA: EV2. The coefficients have been grouped together based on their sign.}
\centering
\begin{tabular}{l l l}

\hline\hline
-                       & Sample     & + \\ \hline
R4570 -0.60              & All (25\%) & $FWHM$(H$\beta$) 0.62 \\
$F_{\text{O}}$ -0.42     &            &                     \\
$F$([OIII]) -0.27        &            &                     \\
density -0.12            &            &                     \\
& & \\
$F_{\text{O}}$ -0.53     & RD (23\%)      & $FWHM$(H$\beta$) 0.51       \\
R4570 -0.43              &                       & $F_{\text{R}}$ 0.37  \\
$F$([OIII]) -0.35        &                       & density 0.14  \\
& & \\
R4570 -0.70              & XD (23\%) & $FWHM$(H$\beta$) 0.67 \\
$F_{\text{X-ray}}$ -0.21 &                       & $F$([OIII]) 0.08 \\ 
density -0.10            &                       &  \\
$F_{\text{O}}$ -0.05     &                       & \\ \hline

\end{tabular}
\label{tab:lrgpcaev2}
\end{minipage}
\end{table}

{\bf EV3} explains 17\% -- 20\% of the variance. Density has the strongest impact on EV3 in all samples. All the other parameters are negligible, except $F_{\text{R}}$ , which correlates with the density in the RD sample. This supports the results obtained in Section~\ref{sec:density} inferring that the large-scale environment and radio properties are connected. Otherwise the fact that EV3 is completely dominated by the large-scale environment parameter, with the other variables affecting very little, further indicates that the large-scale environment density does not appear to be connected to the intrinsic properties of NLS1 galaxies, at least with this sample and set of parameters.

\begin{table}[ht]
\begin{minipage}{0.5\textwidth}
\caption{Results of the LRG PCA: EV3. The coefficients have been grouped together based on their sign.}
\centering
\begin{tabular}{l l l}

\hline\hline
-                     & Sample     & + \\ \hline
$F_{\text{O}}$ -0.09  & All (20\%) & density 0.98         \\ 
$F$([OIII]) -0.07     &            & $FWHM$(H$\beta$) 0.14 \\
                      &            & R4570 0.05         \\
& & \\
$FWHM$(H$\beta$) -0.29 & RD (17\%)    & density 0.68  \\
$F$([OIII]) -0.04      &                     & $F_{\text{R}}$ 0.57 \\
                       &                     & R4570 0.36     \\
                       &                     & $F_{\text{O}}$ 0.03 \\
& & \\
$F_{\text{X-ray}}$ -0.11  & XD (17\%)  & density 0.99 \\
$F_{\text{O}}$ -0.11      &                   & $FWHM$(H$\beta$) 0.06 \\
R4570 -0.04               &                   & $F$([OIII]) 0.03 \\ \hline
\end{tabular}
\label{tab:lrgpcaev3}
\end{minipage}
\end{table}

\subsubsection{BOSS PCA}
\label{sec:bosspca}

\begin{figure*}[ht!]
\centering
\includegraphics[width=0.99\textwidth]{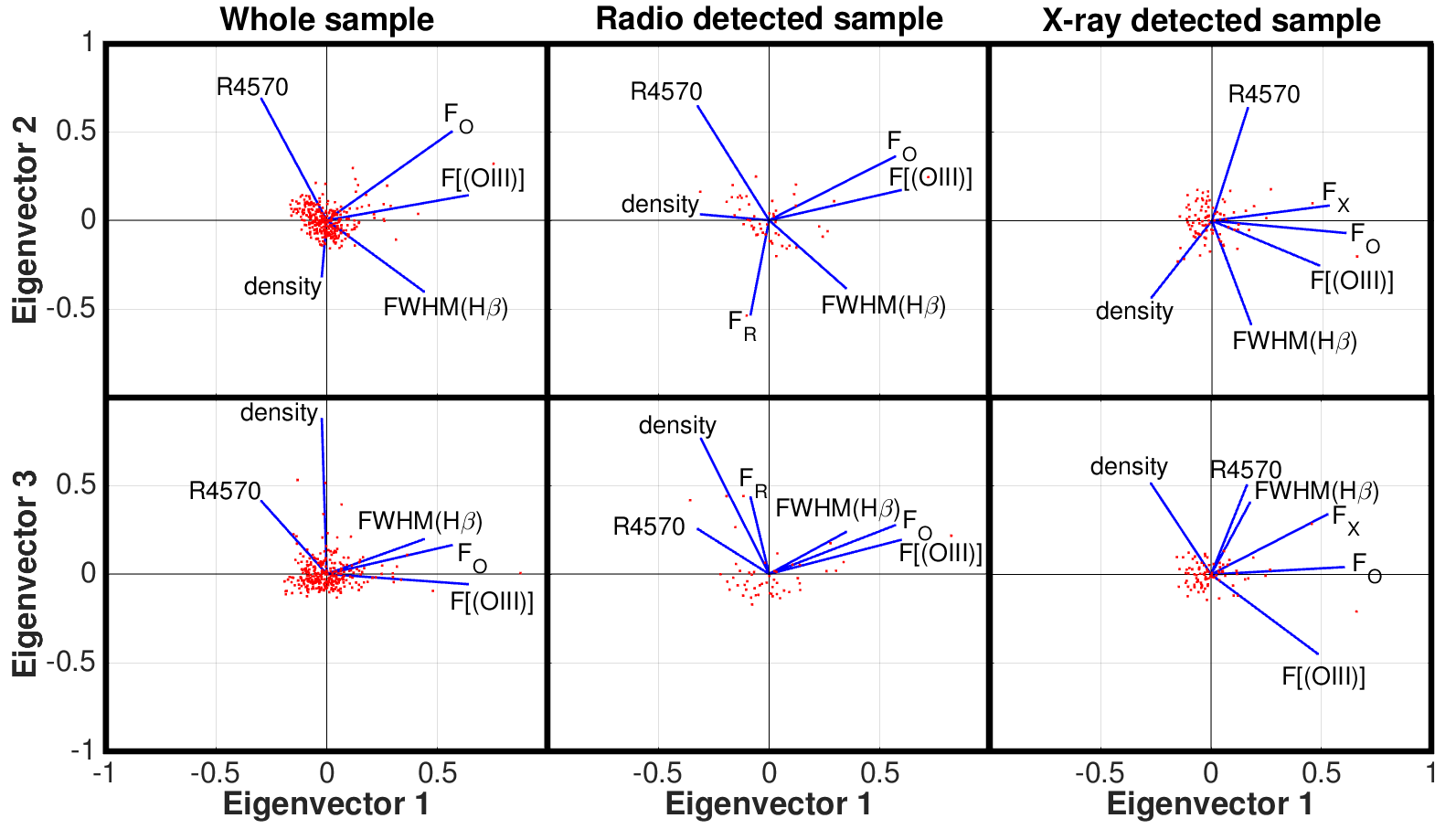}
\caption{Biplots of EVs 1, 2 and 3, and variables of the BOSS PCA.}
\label{fig:biplotboss}
\end{figure*}

For the sources with the BOSS LDF data, we performed the PCA as described in ~\ref{sec:lrgpca}, substituting the LRG LDF data with the BOSS LDF data. The whole sample consists of 312 sources, the RD sample of 51 sources, and the XD sample of 91 sources. The PCA coefficients for the first three principal components are shown in Tables~\ref{tab:bosspcaev1},~\ref{tab:bosspcaev2}, and ~\ref{tab:bosspcaev3}, and the biplots are shown in Figure~\ref{fig:biplotboss}.

{\bf EV1} 28\% -- 38\% of the variance is explained by EV1. It is similar to the EV1 of the LRG PCA; in every sample $F$([O III]) and $F_{\text{O}}$ are the strongest contributors. They are anticorrelated with R4570 in the whole and RD samples, and correlated with $F_{\text{X-ray}}$ in the XD sample. Overall the R4570 contribution is weaker than in LRG PCA EV1. In the RD sample, the contribution of the density is about the same as of R4570, but neither is particularly strong.

\begin{table}[ht]
\begin{minipage}{0.5\textwidth}
\caption{Results of the BOSS PCA: EV1. The coefficients have been grouped together based on their sign.}
\centering
\begin{tabular}{l l l}

\hline\hline
-             & Sample     & + \\ \hline
R4570 -0.29   & All (34\%) & $F$([OIII]) 0.64      \\ 
density -0.02 &            & $F_{\text{O}}$ 0.56    \\
              &            & $FWHM$(H$\beta$) 0.44 \\
 & & \\
R4570 -0.32          & RD (28\%) & $F$([OIII]) 0.59 \\
density -0.31        &                  & $F_{\text{O}}$ 0.57 \\
$F_{\text{R}}$ -0.08 &                  & $FWHM$(H$\beta$) 0.35 \\
& & \\
density -0.27  & XD (38\%)     & $F_{\text{O}}$ 0.60  \\
               &                      & $F_{\text{X-ray}}$ 0.53 \\
               &                      & $F$([OIII]) 0.48 \\ 
               &                      & $FWHM$(H$\beta$) 0.18 \\
               &                      & R4570 0.16  \\ \hline

\end{tabular}
\label{tab:bosspcaev1}
\end{minipage}
\end{table}

{\bf EV2} accounts for 21\% -- 25\% of the variance. The overall trend is that EV2 is led by the anticorrelation between R4570 and $FWHM$(H$\beta$), suggesting that it is similar to the traditional EV1 as in the LRG PCA. In addition to this anticorrelation, $F_{\text{O}}$ correlates with R4570 in the whole and RD samples. $F_{\text{R}}$ has a strong contribution in the RD sample, and the density parameter comes forth in the XD sample.

\begin{table}[ht]
\begin{minipage}{0.5\textwidth}
\caption{Results of the BOSS PCA: EV2. The coefficients have been grouped together based on their sign.}
\centering
\begin{tabular}{l l l}

\hline\hline
-                    & Sample     & + \\ \hline
R4570 -0.69          & All (23\%) & $FWHM$(H$\beta$) 0.40 \\ 
$F_{\text{O}}$ -0.50 &            & density 0.32       \\
$F$([OIII]) -0.14    &            &                     \\
& & \\
R4570 -0.64          & RD (21\%)      & $F_{\text{R}}$ 0.53        \\
$F_{\text{O}}$ -0.36 &                       & $FWHM$(H$\beta$) 0.38   \\
$F$([OIII]) -0.17    &                       &  \\
density -0.03        &                       &   \\
& & \\
R4570 -0.63              & XD (25\%) & $FWHM$(H$\beta$) 0.58  \\
$F_{\text{X-ray}}$ -0.09 &                  & density 0.43 \\ 
                         &                  & $F$([OIII]) 0.25 \\
                         &                  & $F_{\text{O}}$ 0.07 \\ \hline

\end{tabular}
\label{tab:bosspcaev2}
\end{minipage}
\end{table} 

{\bf EV3} 14\% -- 21\% of the variance is explained by EV3. In the whole and RD samples the large-scale environment density is the sole strong contributor. EV3 of the XD sample is slightly different and distinguishes between $F$([O III]), density, and R4570.

\begin{table}[ht]
\begin{minipage}{0.5\textwidth}
\caption{Results of the BOSS PCA: EV3. The coefficients have been grouped together based on their sign.}
\centering
\begin{tabular}{l l l}

\hline\hline
-                & Sample     & + \\ \hline
$F$([OIII]) -0.06  & All (21\%) & density 0.87    \\ 
                 &            & R4570 0.41  \\
                 &            & $FWHM$(H$\beta$) 0.20 \\
                 &            & $F_{\text{O}}$ 0.16       \\
& & \\
         & RD (17\%)  & density 0.76 \\
         &                   & $F_{\text{R}}$ 0.43 \\
         &                   & $F_{\text{O}}$ 0.27   \\
         &                   & R4570 0.25     \\
         &                   & $FWHM$(H$\beta$) 0.24   \\
         &                   & $F$([OIII]) 0.19  \\
& & \\
$F$([OIII]) -0.45 & XD (14\%)& density 0.51 \\
                &                      & R4570 0.50 \\
                &                      & $FWHM$(H$\beta$) 0.41 \\ 
                &                      & $F_{\text{X-ray}}$ 0.34 \\ 
                &                      & $F_{\text{O}}$ 0.04 \\ \hline
\end{tabular}
\label{tab:bosspcaev3}
\end{minipage}
\end{table} 

\subsubsection{EV correlations}
\label{sec:evcorr}

In order to gain more insight into the connection between the EVs and the physical properties of our sources, we calculated the Spearman rank correlation coefficients between EVs of both LRG and BOSS PCA, and selected intrinsic AGN properties --  Eddington ratio ($L_{\text{bol}}$ / $L_{\text{Edd}}$), $M_{\text{BH}}$, and $\lambda$ $L_{5100}$ -- for the whole, RD, and XD samples. To compute the Eddington ratio we used the estimations $L_{\text{bol}}$ = 9$\lambda$ $L_{5100}$ \citep{2000kaspi1} and $L_{\text{Edd}}$ = 1.3$\times 10^{38}$ $M_{\text{BH}}$ / $M_{\sun}$ \citep{2012xu1}. Logarithmic mean, minimum, and maximum values for the Eddington ratio in the whole sample are -0.15, -0.88, and 1.20, respectively. The correlation results are shown in Tables~\ref{tab:evcorr1} and~\ref{tab:evcorr2}.

{\bf LRG} EV1 correlates with the Eddington ratio to some extent in the whole and RD samples, but not in the XD sample. There is a weak correlation with the black hole mass in all samples. EV2 of our sample is closest to the `traditional EV1', and correlates strongly with the Eddington ratio in all samples, and significantly, but not so strongly, with the black hole mass. EV3 does not correlate strongly with any of the properties, which is understandable since the large-scale environment does not directly affect any singular properties, but rather the overall evolution of the galaxies. The only exception to this is the RD sample in which EV3 shows some correlation with the Eddington ratio.

\begin{table*}[ht]
\caption{Spearman rank correlation between EVs and the intrinsic properties (probability values in parentheses):
whole, RD and XD samples with SDSS LRG LDF data. Correlations in boldface have p$<$0.05.}
\centering
\begin{tabular}{l l l l l}
\hline\hline
    &     & log $L / L_{\text{Edd}}$ & log $M_{\text{BH}}$  & log $\lambda L_{5100}$\\ \hline
EV1 & all & {\bf -0.506} (0.000)     & {\bf 0.291} (0.000)  & {\bf -0.083} (0.011) \\
    & RD  & {\bf -0.452} (0.000)     & {\bf 0.360} (0.000)  & 0.030 (0.724) \\
    & XD  & -0.080 (0.109)           & {\bf 0.209} (0.000)  & {\bf 0.118} (0.017) \\
EV2 & all & {\bf -0.764} (0.000)     & {\bf 0.394} (0.000)  & {\bf -0.112} (0.001) \\
    & RD  & {\bf 0.717} (0.000)      & {\bf -0.314} (0.000) & 0.138 (0.107) \\
    & XD  & {\bf -0.827} (0.000)     & {\bf 0.453} (0.000)  & {\bf -0.108} (0.029) \\
EV3 & all & {\bf -0.157} (0.000)     & {\bf 0.100} (0.002)  & -0.003 (0.924)\\ 
    & RD  & {\bf -0.419} (0.000)     & {\bf 0.294} (0.001)  & 0.015 (0.859)\\
    & XD  & {\bf -0.140} (0.005)     & -0.042 (0.394)       & {\bf -0.141} (0.005) \\ \hline
    
\end{tabular}
\label{tab:evcorr1}
\end{table*}

{\bf BOSS} EV1 correlates strongly with $M_{\text{BH}}$ and is clearly comparable to EV2 of previous studies. There is a slight correlation with the continuum luminosity, $\lambda$ $L_{5100}$, especially in the XD sample. In addition`, the whole sample correlates with the Eddington ratio to some extent. The correlation of EV2 with the Eddington ratio is strong in all samples, indicating that it is the main driver behind this EV. EV2 is clearly the traditional AGN EV1. EV3 correlations are rather ambiguous, which can be explained by the strong contribution of the density -- which does not correlate with any other properties -- to EV3. The weak correlations seen between EV3 and the intrinsic properties are probably induced by other parameters and not the density.

\begin{table*}[ht]
\caption{Spearman rank correlation between EVs and the intrinsic properties (probability values in parentheses):
whole, RD and XD samples with SDSS BOSS LDF data. Correlations in boldface have p$<$0.05.}
\centering
\begin{tabular}{l l l l l}
\hline\hline
    &     & log $L / L_{\text{Edd}}$ & log $M_{\text{BH}}$  & log $\lambda L_{5100}$\\ \hline
EV1 & all & {\bf -0.407} (0.000)     & {\bf 0.749} (0.000)  & {\bf 0.487} (0.000) \\
    & RD  & -0.246 (0.082)           & {\bf 0.639} (0.000)  & {\bf 0.432} (0.001) \\
    & XD  & -0.100    (0.362)        & {\bf 0.666} (0.000)  & {\bf 0.769} (0.000) \\
EV2 & all & {\bf -0.709} (0.000)     & 0.061 (0.286)        & {\bf -0.412} (0.000) \\
    & RD  & {\bf -0.732} (0.000)     & 0.059 (0.678)        & {\bf -0.372} (0.007) \\
    & XD  & {\bf -0.701} (0.000)     & {\bf 0.436} (0.000)  & -0.017 (0.874)\\
EV3 & all & -0.060  (0.294)          & {\bf 0.329} (0.000)  & {\bf 0.307} (0.000) \\ 
    & RD  & 0.051  (0.721)           & {\bf 0.517} (0.000)  & {\bf 0.559} (0.000)\\
    & XD  & {\bf -0.354} (0.001)     & {\bf 0.359} (0.001)  & 0.189 (0.074)\\ \hline
    
\end{tabular}
\label{tab:evcorr2}
\end{table*}

\section{Discussion}
\label{sec:disc}

\subsection{NLS1 galaxies in the AGN family}

NLS1 galaxies have been found to be morphologically young, late-type sources in previous studies concentrating on smaller samples. Our findings support this scenario. In the SDSS Main Galaxy sample, 23\% of NLS1 sources reside in groups. This number is low compared to galaxies in general; in the flux-limited SDSS Main Galaxy sample 48\% of galaxies are in groups \citep{2014A&A...566A...1T}, and according to \citet{1996bahcall1} 60\% of galaxies are in groups or clusters, leaving 40\% of galaxies to be in pairs or field galaxies. The excess of field NLS1 galaxies is consistent with their young age. Also in the framework of the density -- morphology relation \citep{1931hubble1,1980dressler1} -- more evolved galaxies are found in denser regions -- our study confirms the young nature of the NLS1 population. The average total mass of the groups with 4 -- 15 members in \citet{2014A&A...566A...1T} is $\sim$ 10$^{13} h^{-1} M_{\sun}$, which is of the same order as for more luminous quasars \citep{2009mandelbaum1,2013koutoulidis1,2013fanidakis1,2013shen1} and approximately the same as for the groups our NLS1 galaxies reside in.
 
Our results clearly distinguish between NLS1 and BLS1 sources in terms of average densities. This discrepancy contradicts the simple orientation-based unification model of NLS1 and BLS1 galaxies in which, when assuming a disk-like BLR, the observed differences can be explained as orientation effects, and NLS1 sources can be unified with BLS1 sources \citep{2008decarli1,2017rakshit1}. If BLS1 galaxies were actually the parent population of NLS1 galaxies, we would expect their spatial distributions to be similar. Our findings also contradict the results in \citet{2014ermash1} who used SDSS DR7 to study the spatial densities of NLS1 and BLS1 galaxies as functions of the large-scale galaxy density. They found that the ratio $N_{NLS1}/N_{BLS1}$ is constant and does not depend on large-scale galaxy density. We do not find the fraction to be constant.

However, BLS1 galaxies are diverse in their properties; for example, their bulges can be classical, composite, or pseudobulges, whereas the bulges in the majority of NLS1 galaxies are pseudobulges. This diversity has not been taken into account in most papers studying, and finding, differences between NLS1 and BLS1 sources (e.g. \citet{2003crenshaw1, 2006deo1, 2010sani1}), nor was it taken into account in \citet{2011lietzen1}. It is, therefore, possible that a subset of BLS1 galaxies is part of the parent population of NLS1 galaxies.

A similar challenge due to heterogeneous samples has been encountered before; \citet{2003tran1} suggested that studies of the differences between Sy1 and Sy2 galaxies should be reanalysed taking into account the fact that Sy2 galaxies are a heterogeneous class consisting of Sy2 sources with powerful hidden Sy1 nuclei and of `pure' Sy2 sources with weak nuclei and very weak or nonexistent BLR \citep{1997heisler1, 2001tran1, 2003tran1, 2012marinucci1}. They argue that, for example, \citet{2001schmitt1} did not find differences because due to selection effects they compare Sy1 and hidden broad-line region (HBLR) Sy2 galaxies, which by definition should be similar. They reanalysed some of the studies and found that when Sy2 sources are divided into two populations, the results change, revealing Sy1 and HBLR Sy2 sources to be similar, and non-HBLR Sy2 sources to be different. They suggest an evolutionary link between pure Sy2 galaxies, and HBLR Sy2 and Sy1 galaxies; either non-HBLR Sy2 sources evolve to more powerful HBLR Sy2 and Sy1 sources once the AGN activity is triggered (e.g. \citet{2000nicastro1,2011wu1}), or non-HBLR Sy2 galaxies have exhausted their fuel and returned to a dormant state \citep{2011yu1, 2013yu1}. A similar evolutionary scenario has been proposed for NLS1 and BLS1 galaxies; assuming that the narrow broad lines are due to the undermassive BH, the fast growing NLS1 galaxies would eventually evolve to Sy1 galaxies with higher-mass black holes and thus broader broad lines, that is, BLS1 galaxies. Our results favour this idea over the unification by orientation scenario.

Moreover, a few higher-inclination steep-spectrum NLS1 sources with extended radio emission have recently been found \citep{2015doi1,2017congiu1}, proving that in these sources the narrowness of the broad lines is due to the low black hole mass and not orientation effects, supporting the scenario that no additional `misaligned' broad-line Type 1 parent population is necessarily needed. However, the parent population studies are still in the early phase and, for example, edge-on compact steep-spectrum (CSS) sources and disk-hosted radio galaxies have been proposed `to be possible Type 1 parent populations of jetted NLS1 sources \citep{2015berton1,2016berton1, 2017berton1}. The nature of the obscured, or Type 2, parent population remains unclear; for the jetted sources it might include narrow-line CSS sources and narrow-line disk-hosted radio galaxies. \citet{2006zhang1} suggested that non-HBLR Sy2 galaxies could be the Type 2 counterparts of NLS1 galaxies and unified based on orientation. This result, however, remains debatable since the sources have many unexplained differences, for example, non-HBLR Sy2 galaxies do not show Fe II emission lines \citep{2013yu1}. 

\begin{figure*}[ht!]
\centering
\includegraphics[width=0.99\textwidth]{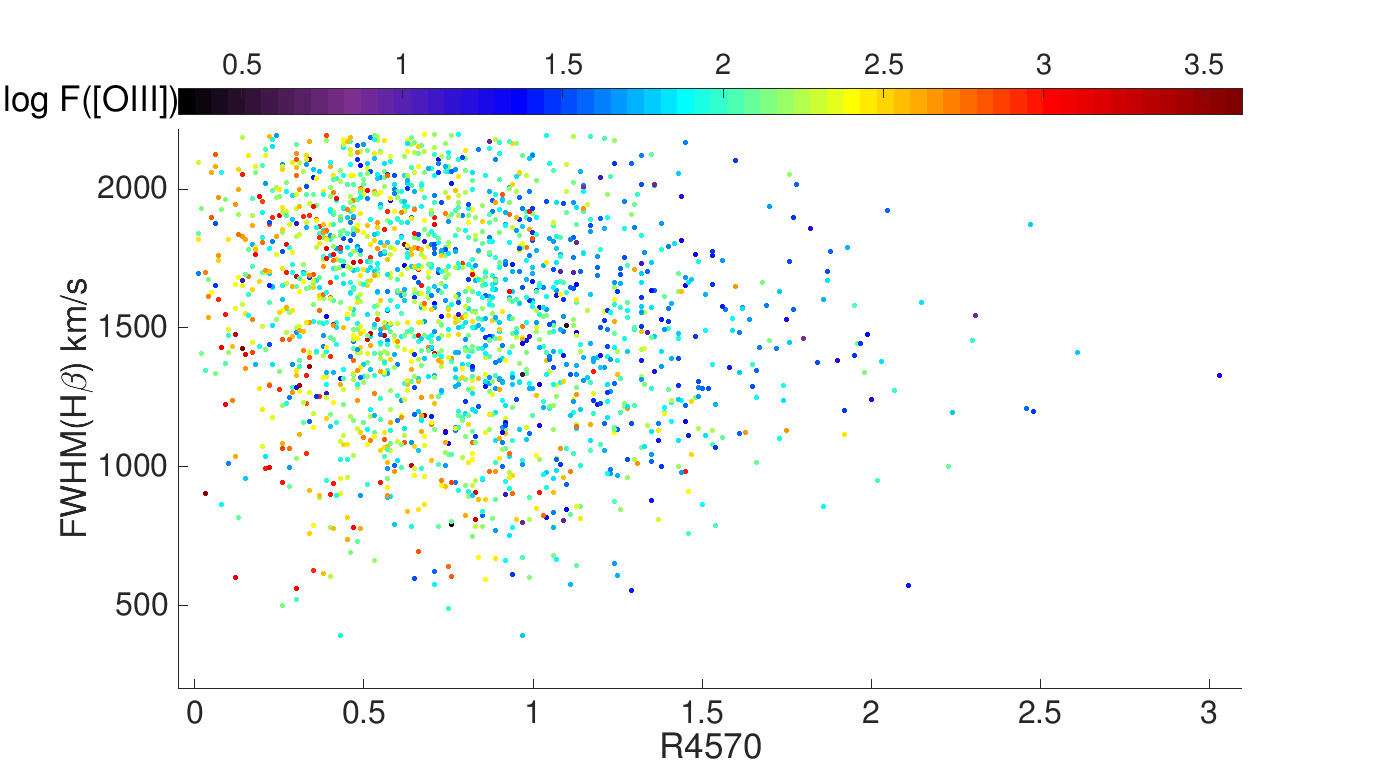}
\caption{Narrow $FWHM$(H$\beta$) extreme of the 4DE1 continuum, populated by NLS1 galaxies which follow the same trends as AGN in general. Abscissa is the strength of Fe II compared to the broad H$\beta$, and ordinate is the $FWHM$(H$\beta$). The flux density of [O III] is shown in colour; bluer colour indicates lower flux density and redder colour higher flux density.}
\label{fig:3Dcorrspace}
\end{figure*}

Our analysis based on PCA supports the idea that NLS1 sources are a part of the continuous AGN spectrum; EVs 1 and 2 found in this study are a manifestation of the 4DE1 AGN correlation space -- in our case without C IV -- dominated by the interplay between $FWHM$(H$\beta$), R4570, and [O III]. These three parameters were studied in more detail, for example in \citet{2014shen1}. They suggest that the variations in $FWHM$(H$\beta$), R4570, and [O III], and the correlations between them, can be explained based only on the Eddington ratio and the orientation. The increasing Eddington ratio leads to increasing R4570 and decreasing [O III] strength, and the scatter in $FWHM$(H$\beta$) at a fixed R4570 is mainly due to the orientation of a flattened BLR, and not the black hole mass and virialised BLR. Their sample consists mainly of broad-line AGN. NLS1 galaxies represent the narrow $FWHM$(H$\beta$), strong Fe II, and weak [O III] extreme of this continuum. In Fig.~\ref{fig:3Dcorrspace} we show a plot similar to Fig.~1 in \citet{2014shen1}. NLS1 sources follow the same trend as broad-line AGN and complete one extreme of the continuum; the flux density of [O III] decreases as the Fe II strength increases, and there does not seem to exist any vertical trends correlations with the $FWHM$(H$\beta$). The same domination of $FWHM$(H$\beta$), R4570 and [O III] in the PCA can be seen in Figures~\ref{fig:biplotlrg} and~\ref{fig:biplotboss}. We cannot, however, confirm the role of the orientation as the driver of the scatter in $FWHM$(H$\beta$) values since our black hole mass estimate is orientation-dependent. If the $FWHM$(H$\beta$) depended only on the orientation then BLS1 sources should be the parent population of NLS1 galaxies, which, based on our results, is improbable.

\subsection{NLS1 galaxies as a class}

In the SDSS Main Galaxy sample only 25\% of RD and 34\% of RS NLS1 galaxies reside in groups. The average richness of the groups is similar, but RD sources reside in physically slightly smaller, and thus denser, groups. This indicates that not the group membership itself but the density of the group is correlated with the radio properties of NLS1 galaxies.

The density of the large-scale environment has the capacity of transfiguring the radio characteristics of NLS1 sources, probably via interactions or merging; the incidence of RL sources increases with the increasing large-scale environment density. However, the average black hole masses of the subsamples -- RS, RQ, and RL -- within a certain redshift range are almost similar. This suggests that the environment is indeed causing the differences in the radio loudness. Radio loudness does not correlate with the black hole mass, or, at a given redshift; NLS1 sources with more massive black holes are not more probable to be radio loud. However, on average the black hole mass in RL sources is higher because they are farther away; there is a clear excess of RL and deficit of RQ sources at high redshift. This fits well with the hierarchical evolution model of AGN, discussed in ~Section \ref{sec:envdata}.

The average black hole masses within the different density regions are similar. For example, in SDSS LRG LDF, RL sources have similar average black hole masses in voids, intermediate density regions, and superclusters. The same holds for RQ and RS sources. This might suggest that, at least up to this point, the accretion history of these sources has been similar in all density regions.

The distribution of sources to voids, intermediate density regions, and superclusters changes with increasing radio loudness; RL sources have a higher probability of being found in superclusters. An exception to this are RS sources. The biggest fraction of RS sources resides in voids; these are probably the `true' RS NLS1 galaxies. However, in superclusters, the fraction is clearly higher than for RQ sources. This might indicate, and be explained by the misclassification of a fraction of RS sources, meaning that the subsample is mixed. The two-sample KS test indeed suggests that the density distributions of our subsamples -- RS versus RD, RQ versus RL, and RS versus RL -- are drawn from the same continuous distribution, which further indicates that
the subsamples are not well defined.

Misclassifications and thus heterogeneous samples are one of the biggest caveats in extensive AGN population studies, and one of the reasons why large statistical studies of NLS1 galaxies and also other AGN classes give inconsistent results. In our study there are a few possible reasons for misclassifications. Data are incomplete, even though optical data cannot be incomplete since NLS1 sources are selected based on that, but we might lose some weak radio emitters from the RQ sample to the RS sample. Due to enhanced star formation, the RL subsample probably includes a few RQ and RS sources, and the RQ subsample might include RS sources. Also the radio loudness value is time-dependent, because NLS1 sources are variable in optical and radio bands (e.g. \citet{2017lahteenmaki1}, Gab\'{a}nyi et al. in prep.) Moreover, there is clear evidence that some RS sources have jets and are thus severely misclassified (L\"{a}hteenm\"{a}ki et al. in prep.).

PCA results of the whole, RD, and XD samples are mostly consistent. In the majority of cases the first two EVs are dominated by the 4DE1 components; R4570, $FWHM$(H$\beta$), and F([O III]), and the third EV by the density parameter. The appearance of the `traditional' EV1, R4570 -- $FWHM$(H$\beta$), only as the second most important principal component is probably due to our pure NLS1 sample; the variance of $FWHM$(H$\beta$) is small compared to mixed samples, but the variance in R4570 and F([O III]) is of comparable magnitude. It has been suggested that the anticorrelation of R4570 and F([O III]) is driven by the Eddington ratio, and the variance in $FWHM$(H$\beta$) is only due to the orientation \citep{2014shen1}. Interestingly, in our sample, the Eddington ratio 
correlates poorly with EV1 that is dominated by the R4570 -- F([O III]) anticorrelation. Instead, it correlates very well with EV2, dominated by R4570 and $FWHM$(H$\beta$). To some extent this might be because $FWHM$(H$\beta$) is used to estimate the Eddington ratio.

Adding $F_{\text{R}}$ or $F_{\text{X-ray}}$ to PCA changes the EVs slightly. Radio flux density emerges in EVs 2 and 3, correlating with $FWHM$H($\beta$) and the density parameter, respectively. Correlation of the radio flux density and $FWHM$H($\beta$) supports the hypothesis that radio brighter-, and also radio louder- (since radio loudness correlates with the radio flux density) sources have on average broader $FWHM$(H$\beta$). $FWHM$(H$\beta$) does not evolve with the redshift so this correlation is not induced by an increasing fraction of RL sources being farther away. Correlation of $F_{\text{R}}$ and the density support the results from the average density calculations of the different subsamples. In the XD sample $F_{\text{X-ray}}$ and $F_{\text{O}}$ emerge in the first EV. This might be induced by the initial correlation of $F_{\text{X-ray}}$ and $F_{\text{O}}$ , which then enhances variance in their direction.

Overall the density parameter appears to have more impact on BOSS PCA than on LRG PCA. BOSS LDF has a smaller smoothing scale (8 $h^{-1}$ Mpc) than LRG LDF (16 $h^{-1}$ Mpc) which means that it traces the structures at a smaller scale, which might induce the difference in PCA. Another possibility is that the LRG and BOSS samples are intrinsically slightly different; the excess of RL sources at higher redshifts might cause the density to appear more important in the PCA.

\subsection{From RL to jetted NLS1 galaxies}
\label{sec:starformation}

Using the more or less arbitrarily defined value of radio loudness as a dividing parameter between NLS1 subclasses results in poorly defined, mixed samples. This affects the results. The subsamples should be divided based on their real physical properties, for example, jetted and non-jetted sources, as proposed in \citet{2016padovani1}. This would require extensive studies of individual sources, which in the case of extensive samples is not viable, and the radio loudness parameter must thus be used as a proxy of the nuclear activity. However, we can begin to address this issue by studying the environments of those sources that are known to have jets.

For this we selected a sample of NLS1 galaxies whose radio emission should be dominated by non-thermal emission from the AGN.
We included all gamma-ray detected NLS1 galaxies and NLS1 sources which have kpc- or pc-scale jets. In addition we included sources detected at the Mets\"{a}hovi Radio Observatory at 37~GHz(\citet{2017lahteenmaki1} and L\"{a}hteenm\"{a}ki et al. in prep), and sources in which, based on the q22 parameter \citep{2015caccianiga1}, the radio emission is dominated by the jet emission.

The q22 parameter can be used to distinguish sources in which the radio loudness is due to the jet from sources in which it might be induced by enhanced star formation \citep{2015caccianiga1}. It is defined as q22 = log (F$_{22 \mu \text{m}}$ / F$_{1.4 \text{GHz}}$), and based on the fact that jet-dominated and star formation-dominated sources have different radio-to-mid-infrared flux ratios. In sources with $q22<-0.8$ the jet is probably the main source of the radio emission; in sources with $q22>1$ the star formation may contribute significantly to the observed radio emission, especially in sources which have excess mid-infrared emission, that is, W3---W4 $>$ 2.5. Sources in between, with $-0.8<q22<1$, have probably both the star formation and the jet contributing to the radio emission.

Altogether, we have 29 sources that supposedly have a jet and for which we have the large-scale environment data; 17 of those lie in the LRG LDF and 12 in the BOSS LDF. The Mets\"{a}hovi sample includes sources that were initially selected for the observing programme due to their dense environments; two of these have been detected and might considerably increase the average density shown below. The average densities for the jetted NLS1 galaxies in the LRG sample, with and without the two density-selected sources, and in the BOSS sample are presented in Table~\ref{tab:jetnls1}. For comparison, the average densities for the mixed and star formation-dominated subsamples are presented.

\begin{table}[ht]
\caption[]{Average large-scale environment densities of jetted, mixed emission, and non-jetted subsamples of NLS1 galaxies.}
\centering
\begin{tabular}{l l l}
\hline\hline
                                  & N     & Average density \\ \hline
jetted, LRG                       & 17    & 2.48 $\pm$ 0.57  \\
jetted, LRG, w/o density-selected & 15    & 1.75 $\pm$ 0.31  \\
mixed, LRG                        & 82    & 1.68 $\pm$ 0.15  \\
star formation, LRG               & 57    & 1.44 $\pm$ 0.14  \\  \hline
jetted, BOSS                      & 12    & 2.05 $\pm$ 0.81  \\ \hline

\end{tabular}
\label{tab:jetnls1}
\end{table}

The subsample sizes are relatively small and thus the results are not unambiguous, however, the average density of the jetted NLS1 sources in the LRG LDF is almost similar to the traditionally defined RL NLS1 sources (1.71$\pm$0.14). Compared to mixed, and especially the star formation dominated sources, the jetted sources are more likely to reside in denser environments. This result is consistent with the results of the various RL subsamples, since it is probable that the increasing radio loudness increases the probability that the radio emission originates in the jet. It is noteworthy that the average density of the star formation-dominated subsample is similar to that of the RS subsample. It is understandable since in these sources the contribution of the AGN to the radio emission might be negligible or non-existent, and the AGN in them is probably radio quiet or silent. This result does not exclude the possibility that jetted sources could have enhanced star formation, and thus does not tell us anything about how the sources with different amounts of star formation are distributed to voids, intermediate density regions, and superclusters. 

It is not surprising that NLS1 galaxies reside in less dense supercluster-scale environments when compared to the AGN samples in \citet{2011lietzen1}; most of their samples consist of source types that are considered to be evolved, whereas the majority of NLS1 galaxies are thought to be young, unevolved spiral galaxies. The result supports the hierarchical galaxy evolution model in which the more evolved galaxies, and also AGN, are found in denser environments. The pronounced difference between the environments of the jetted NLS1 galaxies and other jetted AGN classes studied in \citet{2011lietzen1} is particularly interesting. Their samples included BL Lac objects (2.50$\pm$0.20), flat-spectrum radio galaxies (2.60$\pm$0.07), FR I radio galaxies (3.01$\pm$0.07), and FR II radio galaxies (3.20$\pm$0.04), all of which have significantly higher average large-scale environment densities than the jetted NLS1 galaxies (1.75$\pm$0.31). This confirms that triggering a jet in diverse environments, spanning from voids to superclusters, is possible. Wheras the results indicate that the denser large-scale environment affects the incidence of jets, the triggering mechanism remains unclear. A small percentage (8\%--16\%, \citet{2007ohta1}) of NLS1 galaxies shows signs of interaction or merging \citep[20\%--30\% for Seyfert galaxies in general, ][]{2001schmitt2}, suggesting that secular processes dominate the growth and evolution of NLS1 sources. Extensive studies of the host galaxies of the different, especially the jetted, NLS1 sources are needed to examine whether a link between jets, and interaction and merging exists. The error in the average density of jetted BOSS sources is relatively large, and unfortunately we do not have any comparison samples, but the results are generally consistent with the LRG LDF results.
 
\subsection{Conclusions}
\label{sec:concl}

Our study is the first attempt to study the relationship of the large-scale environment and the intrinsic properties of NLS1 galaxies using a large, statistically significant sample. The large-scale environment is an interesting additional parameter helping us to understand how the changes at the largest cosmic scales emerge in the smaller-scale properties of AGN; for example, their host galaxies and nuclear activity. Our main conclusions are the following.

\begin{itemize}
 \item[$\bullet$] This study supports the young, unevolved nature of NLS galaxies, and shows that the large-scale environment density affects their radio properties; RL, and also jetted NLS1 sources preferably reside in denser regions compared to RQ or non-jetted NLS1 sources.\\
 
  \item[$\bullet$] NLS1 galaxies are a diverse, heterogeneous class of sources which suffers from a considerable amount of misclassification. Our results support the Padovani's view that the traditional division to RL and RQ/RS sources is a severe problem and should be replaced with the division to jetted and non-jetted NLS1 galaxies.\\

 \item[$\bullet$] Orientation is probably not a sufficient explanation for the differences in NLS1 and BLS1 galaxies. NLS1 galaxies are a distinct class of sources, and BLS1 galaxies are not their parent population. However, they could be evolutionarily connected.\\

 \item[$\bullet$] NLS1 sources are located in one extreme of the AGN 4DE1 continuum, but they clearly are distinctive in their nature among the gamma-ray emitting AGN.\\
 
\end{itemize}

Our results indicate that some of the properties of NLS1 galaxies might be affected by the large-scale environment, but it is likely that the large-scale environment has an effect at the scale of the whole galaxy evolution and is not clearly connected to the specific intrinsic properties of NLS1 galaxies, or AGN in general; it affects the long time-scale evolution of galaxies, whereas the nuclear activity of galaxies is variable and intermittent at timescales shorter than galaxy evolution. Since all of the sources in our sample are NLS1 sources and close to one another in the evolutionary sequence, the differences between the sources are not pronounced enough for drawing definite conclusions of the effects of the large-scale environment. A more diverse sample with various AGN classes is needed to study this more effectively. 

In the future, the environment studies should be extended to include smaller-scale environments; for example, the group and cluster scales, which have a bigger contemporary effect on the individual galaxies than the large-scale environment. Another crucial part of the environment research, and closely connected to the local environment, is to study the host galaxy morphologies of an extensive and diverse sample of NLS1 galaxies. Gathering data at all scales of environments will help us to create a bigger picture of the connection of AGN evolution and the environment, and give insight into the triggering mechanism of the jet. Multiwavelength observations, including interferometry, together with environmental data, will be essential in correctly classifying the various NLS1 sources. This is crucial for population-wide studies, including the issues concerning the heterogeneity of the NLS1 class and the intra-class evolution, and will aid in the search of the parent population.

\begin{acknowledgements}

HL is funded by PUT1627 grant from Estonian Research Council. ME was supported by the ETAG project IUT26-2, and by the European Structural Funds grant for the Centre of Excellence `Dark Matter in (Astro)particle Physics and Cosmology' TK133.

This research  made use of the NASA/IPAC Extragalactic Database (NED) which is operated by the Jet Propulsion Laboratory, 
California Institute of Technology, under contract with the National Aeronautics and Space Administration. 

The National Radio Astronomy Observatory is a facility of the National Science Foundation operated under cooperative agreement by Associated Universities, Inc.

This publication makes use of data products from the Wide-field Infrared Survey Explorer, which is a joint project of the University of California,
Los Angeles, and the Jet Propulsion Laboratory/California Institute of Technology, funded by the National Aeronautics and Space Administration.

Funding for the Sloan Digital Sky Survey (SDSS) has been provided by the Alfred P. Sloan Foundation, the Participating Institutions, 
the National Aeronautics and Space Administration, the National Science Foundation, the U.S. Department of Energy, the Japanese Monbukagakusho, 
and the Max Planck Society. The SDSS Web site is http://www.sdss.org/.

The SDSS is managed by the Astrophysical Research Consortium (ARC) for the Participating Institutions. The Participating Institutions are 
The University of Chicago, Fermilab, the Institute for Advanced Study, the Japan Participation Group, The Johns Hopkins University, 
the Korean Scientist Group, Los Alamos National Laboratory, the Max-Planck-Institute for Astronomy (MPIA), the Max-Planck-Institute for Astrophysics (MPA), 
New Mexico State University, University of Pittsburgh, University of Portsmouth, Princeton University, 
the United States Naval Observatory, and the University of Washington.

This research made use of the ROSAT All-Sky Survey data which have been processed at MPE.

\end{acknowledgements}

\bibliographystyle{aa}
\bibliography{artikkeli.bib}

\begin{thebibliography}{115}
\expandafter\ifx\csname natexlab\endcsname\relax\def\natexlab#1{#1}\fi

\bibitem[{{Abazajian} {et~al.}(2009){Abazajian}, {Adelman-McCarthy},
  {Ag{\"u}eros}, {Allam}, {Allende Prieto}, {An}, {Anderson}, {Anderson},
  {Annis}, {Bahcall}, \& et~al.}]{2009abazajian1}
{Abazajian}, K.~N., {Adelman-McCarthy}, J.~K., {Ag{\"u}eros}, M.~A., {et~al.}
  2009, \apjs, 182, 543

\bibitem[{{Abdi} \& {Williams}(2010)}]{2010abdi1}
{Abdi}, H. \& {Williams}, L.~J. 2010, Wiley Interdisciplinary Reviews:
  Computational Statistics, 2

\bibitem[{{Abdo} {et~al.}(2009){Abdo}, {Ackermann}, {Ajello}, {Axelsson},
  {Baldini}, {Ballet}, {Barbiellini}, {Bastieri}, {Battelino}, {Baughman},
  {Bechtol}, {Bellazzini}, {Bloom}, {Bonamente}, {Borgland}, {Bregeon}, {Brez},
  {Brigida}, {Bruel}, {Caliandro}, {Cameron}, {Caraveo}, {Casandjian},
  {Cavazzuti}, {Cecchi}, {Chekhtman}, {Cheung}, {Chiang}, {Ciprini}, {Claus},
  {Cohen-Tanugi}, {Collmar}, {Conrad}, {Costamante}, {Dermer}, {de Angelis},
  {de Palma}, {Digel}, {Silva}, {Drell}, {Dubois}, {Dumora}, {Farnier},
  {Favuzzi}, {Focke}, {Foschini}, {Frailis}, {Fuhrmann}, {Fukazawa}, {Funk},
  {Fusco}, {Gargano}, {Gehrels}, {Germani}, {Giebels}, {Giglietto}, {Giordano},
  {Giroletti}, {Glanzman}, {Grenier}, {Grondin}, {Grove}, {Guillemot},
  {Guiriec}, {Hanabata}, {Harding}, {Hartman}, {Hayashida}, {Hays}, {Hughes},
  {J{\'o}hannesson}, {Johnson}, {Johnson}, {Johnson}, {Kamae}, {Katagiri},
  {Kataoka}, {Kerr}, {Kn{\"o}dlseder}, {Kuehn}, {Kuss}, {Lande}, {Latronico},
  {Lemoine-Goumard}, {Longo}, {Loparco}, {Lott}, {Lovellette}, {Lubrano},
  {Madejski}, {Makeev}, {Max-Moerbeck}, {Mazziotta}, {McConville}, {McEnery},
  {Meurer}, {Michelson}, {Mitthumsiri}, {Mizuno}, {Monte}, {Monzani},
  {Morselli}, {Moskalenko}, {Murgia}, {Nolan}, {Norris}, {Nuss}, {Ohsugi},
  {Omodei}, {Orlando}, {Ormes}, {Paneque}, {Panetta}, {Parent}, {Pavlidou},
  {Pearson}, {Pepe}, {Pesce-Rollins}, {Piron}, {Porter}, {Rain{\`o}}, {Rando},
  {Razzano}, {Readhead}, {Reimer}, {Reimer}, {Reposeur}, {Richards}, {Ritz},
  {Rodriguez}, {Romani}, {Ryde}, {Sadrozinski}, {Sambruna}, {Sanchez},
  {Sander}, {Parkinson}, {Scargle}, {Schalk}, {Sgr{\`o}}, {Smith}, {Spandre},
  {Spinelli}, {Starck}, {Stevenson}, {Strickman}, {Suson}, {Tagliaferri},
  {Takahashi}, {Tanaka}, {Thayer}, {Thompson}, {Tibaldo}, {Tibolla}, {Torres},
  {Tosti}, {Tramacere}, {Uchiyama}, {Usher}, {Vilchez}, {Vitale}, {Waite},
  {Winer}, {Wood}, {Ylinen}, {Zensus}, {Ziegler}, {Fermi/LAT Collaboration},
  {Ghisellini}, {Maraschi}, {Tavecchio}, \& {Angelakis}}]{2009abdo2}
{Abdo}, A.~A., {Ackermann}, M., {Ajello}, M., {et~al.} 2009, \apj, 699, 976

\bibitem[{{Ahn} {et~al.}(2014){Ahn}, {Alexandroff}, {Allende Prieto}, {Anders},
  {Anderson}, {Anderton}, {Andrews}, {Aubourg}, {Bailey}, {Bastien}, \&
  et~al.}]{2014ahn1}
{Ahn}, C.~P., {Alexandroff}, R., {Allende Prieto}, C., {et~al.} 2014, \apjs,
  211, 17

\bibitem[{{Alam} {et~al.}(2015){Alam}, {Albareti}, {Allende Prieto}, {Anders},
  {Anderson}, {Anderton}, {Andrews}, {Armengaud}, {Aubourg}, {Bailey}, \&
  et~al.}]{2015alam1}
{Alam}, S., {Albareti}, F.~D., {Allende Prieto}, C., {et~al.} 2015, \apjs, 219,
  12

\bibitem[{{Ant{\'o}n} {et~al.}(2008){Ant{\'o}n}, {Browne}, \&
  {March{\~a}}}]{2008anton1}
{Ant{\'o}n}, S., {Browne}, I.~W.~A., \& {March{\~a}}, M.~J. 2008, \aap, 490,
  583

\bibitem[{{Bahcall}(1996)}]{1996bahcall1}
{Bahcall}, N.~A. 1996, ArXiv: 9611148

\bibitem[{{Barth} {et~al.}(2008){Barth}, {Bentz}, {Greene}, \&
  {Ho}}]{2008barth1}
{Barth}, A.~J., {Bentz}, M.~C., {Greene}, J.~E., \& {Ho}, L.~C. 2008, \apjl,
  683, L119

\bibitem[{{Beckmann} \& {Shrader}(2012)}]{2012beckmann1}
{Beckmann}, V. \& {Shrader}, C.~R. 2012, {Active Galactic Nuclei}

\bibitem[{{Bentz} {et~al.}(2009){Bentz}, {Peterson}, {Pogge}, \&
  {Vestergaard}}]{2009bentz1}
{Bentz}, M.~C., {Peterson}, B.~M., {Pogge}, R.~W., \& {Vestergaard}, M. 2009,
  \apjl, 694, L166

\bibitem[{{Berton} {et~al.}(2016){Berton}, {Caccianiga}, {Foschini},
  {Peterson}, {Mathur}, {Terreran}, {Ciroi}, {Congiu}, {Cracco}, {Frezzato},
  {La Mura}, \& {Rafanelli}}]{2016berton1}
{Berton}, M., {Caccianiga}, A., {Foschini}, L., {et~al.} 2016, \aap, 591, A98

\bibitem[{{Berton} {et~al.}(2017){Berton}, {Foschini}, {Caccianiga}, {Ciroi},
  {Congiu}, {Cracco}, {Frezzato}, {La Mura}, \& {Rafanelli}}]{2017berton1}
{Berton}, M., {Foschini}, L., {Caccianiga}, A., {et~al.} 2017, ArXiv:1705.07905

\bibitem[{{Berton} {et~al.}(2015{\natexlab{a}}){Berton}, {Foschini},
  {Caccianiga}, {Richards}, {Ciroi}, {Congiu}, {Cracco}, {La Mura},
  {Marafatto}, \& {Rafanelli}}]{2015berton2}
{Berton}, M., {Foschini}, L., {Caccianiga}, A., {et~al.} 2015{\natexlab{a}}, in
  The Many Facets of Extragalactic Radio Surveys: Towards New Scientific
  Challenges, 75

\bibitem[{{Berton} {et~al.}(2015{\natexlab{b}}){Berton}, {Foschini}, {Ciroi},
  {Cracco}, {La Mura}, {Lister}, {Mathur}, {Peterson}, {Richards}, \&
  {Rafanelli}}]{2015berton1}
{Berton}, M., {Foschini}, L., {Ciroi}, S., {et~al.} 2015{\natexlab{b}}, \aap,
  578, A28

\bibitem[{{Blanton} \& {Roweis}(2007)}]{Blanton2007}
{Blanton}, M.~R. \& {Roweis}, S. 2007, \aj, 133, 734

\bibitem[{{Boroson}(2002)}]{2002boroson1}
{Boroson}, T.~A. 2002, \apj, 565, 78

\bibitem[{{Boroson} \& {Green}(1992)}]{1992boroson1}
{Boroson}, T.~A. \& {Green}, R.~F. 1992, \apjs, 80, 109

\bibitem[{{Caccianiga} {et~al.}(2015){Caccianiga}, {Ant{\'o}n}, {Ballo},
  {Foschini}, {Maccacaro}, {Della Ceca}, {Severgnini}, {March{\~a}}, {Mateos},
  \& {Sani}}]{2015caccianiga1}
{Caccianiga}, A., {Ant{\'o}n}, S., {Ballo}, L., {et~al.} 2015, \mnras, 451,
  1795

\bibitem[{{Chen} {et~al.}(2017){Chen}, {Ho}, {Mandelbaum}, {Bahcall},
  {Brownstein}, {Freeman}, {Genovese}, {Schneider}, \& {Wasserman}}]{2017chen1}
{Chen}, Y.-C., {Ho}, S., {Mandelbaum}, R., {et~al.} 2017, \mnras, 466, 1880

\bibitem[{{Cisternas} {et~al.}(2011){Cisternas}, {Jahnke}, {Inskip},
  {Kartaltepe}, {Koekemoer}, {Lisker}, {Robaina}, {Scodeggio}, {Sheth},
  {Trump}, {Andrae}, {Miyaji}, {Lusso}, {Brusa}, {Capak}, {Cappelluti},
  {Civano}, {Ilbert}, {Impey}, {Leauthaud}, {Lilly}, {Salvato}, {Scoville}, \&
  {Taniguchi}}]{2011cisternas1}
{Cisternas}, M., {Jahnke}, K., {Inskip}, K.~J., {et~al.} 2011, \apj, 726, 57

\bibitem[{{Congiu} {et~al.}(2017){Congiu}, {Berton}, {Giroletti}, {Antonucci},
  {Caccianiga}, {Kharb}, {Lister}, {Foschini}, {Ciroi}, {Cracco}, {Frezzato},
  {J{\"a}rvel{\"a}}, {La Mura}, {Richards}, \& {Rafanelli}}]{2017congiu1}
{Congiu}, E., {Berton}, M., {Giroletti}, M., {et~al.} 2017, ArXiv:1704.03881

\bibitem[{{Corbin}(2000)}]{2000corbin1}
{Corbin}, M.~R. 2000, \apjl, 536, L73

\bibitem[{{Crenshaw} {et~al.}(2003){Crenshaw}, {Kraemer}, \&
  {Gabel}}]{2003crenshaw1}
{Crenshaw}, D.~M., {Kraemer}, S.~B., \& {Gabel}, J.~R. 2003, \aj, 126, 1690

\bibitem[{{Decarli} {et~al.}(2008){Decarli}, {Dotti}, {Fontana}, \&
  {Haardt}}]{2008decarli1}
{Decarli}, R., {Dotti}, M., {Fontana}, M., \& {Haardt}, F. 2008, \mnras, 386,
  L15

\bibitem[{{Decarli} {et~al.}(2011){Decarli}, {Dotti}, \&
  {Treves}}]{2011decarli2}
{Decarli}, R., {Dotti}, M., \& {Treves}, A. 2011, \mnras, 413, 39

\bibitem[{{Deo} {et~al.}(2006){Deo}, {Crenshaw}, \& {Kraemer}}]{2006deo1}
{Deo}, R.~P., {Crenshaw}, D.~M., \& {Kraemer}, S.~B. 2006, \aj, 132, 321

\bibitem[{{Doi} {et~al.}(2013){Doi}, {Asada}, {Fujisawa}, {Nagai}, {Hagiwara},
  {Wajima}, \& {Inoue}}]{2013doi1}
{Doi}, A., {Asada}, K., {Fujisawa}, K., {et~al.} 2013, \apj, 765, 69

\bibitem[{{Doi} {et~al.}(2012){Doi}, {Nagira}, {Kawakatu}, {Kino}, {Nagai}, \&
  {Asada}}]{2012doi1}
{Doi}, A., {Nagira}, H., {Kawakatu}, N., {et~al.} 2012, \apj, 760, 41

\bibitem[{{Doi} {et~al.}(2015){Doi}, {Wajima}, {Hagiwara}, \&
  {Inoue}}]{2015doi1}
{Doi}, A., {Wajima}, K., {Hagiwara}, Y., \& {Inoue}, M. 2015, \apjl, 798, L30

\bibitem[{{Dressler}(1980)}]{1980dressler1}
{Dressler}, A. 1980, \apj, 236, 351

\bibitem[{{Ebeling} {et~al.}(2014){Ebeling}, {Stephenson}, \&
  {Edge}}]{2014ebeling1}
{Ebeling}, H., {Stephenson}, L.~N., \& {Edge}, A.~C. 2014, \apjl, 781, L40

\bibitem[{{Einasto} {et~al.}(2014){Einasto}, {Lietzen}, {Tempel}, {Gramann},
  {Liivam{\"a}gi}, \& {Einasto}}]{2014einasto1}
{Einasto}, M., {Lietzen}, H., {Tempel}, E., {et~al.} 2014, \aap, 562, A87

\bibitem[{{Ellison} {et~al.}(2011){Ellison}, {Patton}, {Mendel}, \&
  {Scudder}}]{2011ellison1}
{Ellison}, S.~L., {Patton}, D.~R., {Mendel}, J.~T., \& {Scudder}, J.~M. 2011,
  \mnras, 418, 2043

\bibitem[{{Ermash}(2014)}]{2014ermash1}
{Ermash}, A.~A. 2014, Astronomy Reports, 58, 205

\bibitem[{{Fabian}(2012)}]{2012fabian1}
{Fabian}, A.~C. 2012, \araa, 50, 455

\bibitem[{{Fanidakis} {et~al.}(2013){Fanidakis}, {Macci{\`o}}, {Baugh},
  {Lacey}, \& {Frenk}}]{2013fanidakis1}
{Fanidakis}, N., {Macci{\`o}}, A.~V., {Baugh}, C.~M., {Lacey}, C.~G., \&
  {Frenk}, C.~S. 2013, \mnras, 436, 315

\bibitem[{{Foschini}(2011)}]{2011foschini1}
{Foschini}, L. 2011, in Narrow-Line Seyfert 1 Galaxies and their Place in the
  Universe

\bibitem[{{Foschini} {et~al.}(2015){Foschini}, {Berton}, {Caccianiga}, {Ciroi},
  {Cracco}, {Peterson}, {Angelakis}, {Braito}, {Fuhrmann}, {Gallo}, {Grupe},
  {J{\"a}rvel{\"a}}, {Kaufmann}, {Komossa}, {Kovalev}, {L{\"a}hteenm{\"a}ki},
  {Lisakov}, {Lister}, {Mathur}, {Richards}, {Romano}, {Sievers},
  {Tagliaferri}, {Tammi}, {Tibolla}, {Tornikoski}, {Vercellone}, {La Mura},
  {Maraschi}, \& {Rafanelli}}]{2015foschini1}
{Foschini}, L., {Berton}, M., {Caccianiga}, A., {et~al.} 2015, \aap, 575, A13

\bibitem[{{Gliozzi} {et~al.}(2010){Gliozzi}, {Papadakis}, {Grupe}, {Brinkmann},
  {Raeth}, \& {Kedziora-Chudczer}}]{2010gliozzi1}
{Gliozzi}, M., {Papadakis}, I.~E., {Grupe}, D., {et~al.} 2010, \apj, 717, 1243

\bibitem[{{Goodrich}(1989)}]{1989goodrich1}
{Goodrich}, R.~W. 1989, \apj, 342, 224

\bibitem[{{Greene} \& {Ho}(2005)}]{2005greene1}
{Greene}, J.~E. \& {Ho}, L.~C. 2005, \apj, 630, 122

\bibitem[{{Grupe}(2004)}]{2004grupe1}
{Grupe}, D. 2004, \aj, 127, 1799

\bibitem[{{Gu} {et~al.}(2015){Gu}, {Chen}, {Komossa}, {Yuan}, {Shen}, {Wajima},
  {Zhou}, \& {Zensus}}]{2015gu1}
{Gu}, M., {Chen}, Y., {Komossa}, S., {et~al.} 2015, \apjs, 221, 3

\bibitem[{{Heisler} {et~al.}(1997){Heisler}, {Lumsden}, \&
  {Bailey}}]{1997heisler1}
{Heisler}, C.~A., {Lumsden}, S.~L., \& {Bailey}, J.~A. 1997, \nat, 385, 700

\bibitem[{{Hubble} \& {Humason}(1931)}]{1931hubble1}
{Hubble}, E. \& {Humason}, M.~L. 1931, \apj, 74, 43

\bibitem[{{J{\"a}rvel{\"a}} {et~al.}(2015){J{\"a}rvel{\"a}},
  {L{\"a}hteenm{\"a}ki}, \& {Le{\'o}n-Tavares}}]{2015jarvela1}
{J{\"a}rvel{\"a}}, E., {L{\"a}hteenm{\"a}ki}, A., \& {Le{\'o}n-Tavares}, J.
  2015, \aap, 573, A76

\bibitem[{{Kaspi} {et~al.}(2000){Kaspi}, {Smith}, {Netzer}, {Maoz}, {Jannuzi},
  \& {Giveon}}]{2000kaspi1}
{Kaspi}, S., {Smith}, P.~S., {Netzer}, H., {et~al.} 2000, \apj, 533, 631

\bibitem[{{Khachikian,} \& {Weedman}(1974)}]{1974khachikian1}
{Khachikian,}, E.~Y. \& {Weedman}, D.~W. 1974, \apj, 192, 581

\bibitem[{{King} \& {Pounds}(2015)}]{2015king1}
{King}, A. \& {Pounds}, K. 2015, \araa, 53, 115

\bibitem[{{Kocevski} {et~al.}(2012){Kocevski}, {Faber}, {Mozena}, {Koekemoer},
  {Nandra}, {Rangel}, {Laird}, {Brusa}, {Wuyts}, {Trump}, {Koo}, {Somerville},
  {Bell}, {Lotz}, {Alexander}, {Bournaud}, {Conselice}, {Dahlen}, {Dekel},
  {Donley}, {Dunlop}, {Finoguenov}, {Georgakakis}, {Giavalisco}, {Guo},
  {Grogin}, {Hathi}, {Juneau}, {Kartaltepe}, {Lucas}, {McGrath}, {McIntosh},
  {Mobasher}, {Robaina}, {Rosario}, {Straughn}, {van der Wel}, \&
  {Villforth}}]{2012kocevski1}
{Kocevski}, D.~D., {Faber}, S.~M., {Mozena}, M., {et~al.} 2012, \apj, 744, 148

\bibitem[{{Komossa} {et~al.}(2006){Komossa}, {Voges}, {Xu}, {Mathur}, {Adorf},
  {Lemson}, {Duschl}, \& {Grupe}}]{2006komossa1}
{Komossa}, S., {Voges}, W., {Xu}, D., {et~al.} 2006, \aj, 132, 531

\bibitem[{{Koutoulidis} {et~al.}(2013){Koutoulidis}, {Plionis},
  {Georgantopoulos}, \& {Fanidakis}}]{2013koutoulidis1}
{Koutoulidis}, L., {Plionis}, M., {Georgantopoulos}, I., \& {Fanidakis}, N.
  2013, \mnras, 428, 1382

\bibitem[{{Kuutma} {et~al.}(2017){Kuutma}, {Tamm}, \& {Tempel}}]{2017kuutma1}
{Kuutma}, T., {Tamm}, A., \& {Tempel}, E. 2017, \aap, 600, L6

\bibitem[{{L{\"a}hteenm{\"a}ki} {et~al.}(2017){L{\"a}hteenm{\"a}ki},
  {J{\"a}rvel{\"a}}, {Hovatta}, {Tornikoski}, {Harrison}, {L{\'o}pez-Caniego},
  {Max-Moerbeck}, {Mingaliev}, {Pearson}, {Ramakrishnan}, {Readhead}, {Reeves},
  {Richards}, {Sotnikova}, \& {Tammi}}]{2017lahteenmaki1}
{L{\"a}hteenm{\"a}ki}, A., {J{\"a}rvel{\"a}}, E., {Hovatta}, T., {et~al.} 2017,
  ArXiv: 1703.10365

\bibitem[{{Laor}(2001)}]{2001laor1}
{Laor}, A. 2001, \apj, 553, 677

\bibitem[{{Le{\'o}n Tavares} {et~al.}(2014){Le{\'o}n Tavares}, {Kotilainen},
  {Chavushyan}, {A{\~n}orve}, {Puerari}, {Cruz-Gonz{\'a}lez},
  {Pati{\~n}o-Alvarez}, {Ant{\'o}n}, {Carrami{\~n}ana}, {Carrasco}, {Guichard},
  {Karhunen}, {Olgu{\'{\i}}n-Iglesias}, {Sanghvi}, \&
  {Valdes}}]{2014leontavares1}
{Le{\'o}n Tavares}, J., {Kotilainen}, J., {Chavushyan}, V., {et~al.} 2014,
  \apj, 795, 58

\bibitem[{{Lietzen} {et~al.}(2011){Lietzen}, {Hein{\"a}m{\"a}ki}, {Nurmi},
  {Liivam{\"a}gi}, {Saar}, {Tago}, {Takalo}, \& {Einasto}}]{2011lietzen1}
{Lietzen}, H., {Hein{\"a}m{\"a}ki}, P., {Nurmi}, P., {et~al.} 2011, \aap, 535,
  A21

\bibitem[{{Lietzen} {et~al.}(2012){Lietzen}, {Tempel}, {Hein{\"a}m{\"a}ki},
  {Nurmi}, {Einasto}, \& {Saar}}]{2012lietzen1}
{Lietzen}, H., {Tempel}, E., {Hein{\"a}m{\"a}ki}, P., {et~al.} 2012, \aap, 545,
  A104

\bibitem[{{Lietzen} {et~al.}(2016){Lietzen}, {Tempel}, {Liivam{\"a}gi},
  {Montero-Dorta}, {Einasto}, {Streblyanska}, {Maraston},
  {Rubi{\~n}o-Mart{\'{\i}}n}, \& {Saar}}]{2016lietzen1}
{Lietzen}, H., {Tempel}, E., {Liivam{\"a}gi}, L.~J., {et~al.} 2016, \aap, 588,
  L4

\bibitem[{{Liivam{\"a}gi} {et~al.}(2012){Liivam{\"a}gi}, {Tempel}, \&
  {Saar}}]{2012liivamagi1}
{Liivam{\"a}gi}, L.~J., {Tempel}, E., \& {Saar}, E. 2012, \aap, 539, A80

\bibitem[{{Lister} {et~al.}(2016){Lister}, {Aller}, {Aller}, {Homan},
  {Kellermann}, {Kovalev}, {Pushkarev}, {Richards}, {Ros}, \&
  {Savolainen}}]{2016lister1}
{Lister}, M.~L., {Aller}, M.~F., {Aller}, H.~D., {et~al.} 2016, \aj, 152, 12

\bibitem[{{Mandelbaum} {et~al.}(2009){Mandelbaum}, {Li}, {Kauffmann}, \&
  {White}}]{2009mandelbaum1}
{Mandelbaum}, R., {Li}, C., {Kauffmann}, G., \& {White}, S.~D.~M. 2009, \mnras,
  393, 377

\bibitem[{{Marinucci} {et~al.}(2012){Marinucci}, {Bianchi}, {Nicastro}, {Matt},
  \& {Goulding}}]{2012marinucci1}
{Marinucci}, A., {Bianchi}, S., {Nicastro}, F., {Matt}, G., \& {Goulding},
  A.~D. 2012, \apj, 748, 130

\bibitem[{{Marziani} {et~al.}(2006){Marziani}, {Dultzin-Hacyan}, \&
  {Sulentic}}]{2006marziani1}
{Marziani}, P., {Dultzin-Hacyan}, D., \& {Sulentic}, J.~W. 2006, {Accretion
  onto Supermassive Black Holes in Quasars: Learning from Optical/UV
  Observations}, ed. P.~V. {Kreitler} (Nova Science Publishers), 123

\bibitem[{{Mathur}(2000)}]{2000mathur1}
{Mathur}, S. 2000, \mnras, 314, L17

\bibitem[{{Mathur} {et~al.}(2001){Mathur}, {Kuraszkiewicz}, \&
  {Czerny}}]{2001mathur1}
{Mathur}, S., {Kuraszkiewicz}, J., \& {Czerny}, B. 2001, \na, 6, 321

\bibitem[{{Nicastro}(2000)}]{2000nicastro1}
{Nicastro}, F. 2000, \apjl, 530, L65

\bibitem[{{Ohta} {et~al.}(2007){Ohta}, {Aoki}, {Kawaguchi}, \&
  {Kiuchi}}]{2007ohta1}
{Ohta}, K., {Aoki}, K., {Kawaguchi}, T., \& {Kiuchi}, G. 2007, \apjs, 169, 1

\bibitem[{{Olgu{\'{\i}}n-Iglesias} {et~al.}(2017){Olgu{\'{\i}}n-Iglesias},
  {Kotilainen}, {Le{\'o}n Tavares}, {Chavushyan}, \&
  {A{\~n}orve}}]{2017olguiniglesias1}
{Olgu{\'{\i}}n-Iglesias}, A., {Kotilainen}, J.~K., {Le{\'o}n Tavares}, J.,
  {Chavushyan}, V., \& {A{\~n}orve}, C. 2017, \mnras, 467, 3712

\bibitem[{{Osterbrock}(1978)}]{1978osterbrock1}
{Osterbrock}, D.~E. 1978, Proceedings of the National Academy of Science, 75,
  540

\bibitem[{{Osterbrock}(1981)}]{1981osterbrock1}
{Osterbrock}, D.~E. 1981, \apj, 249, 462

\bibitem[{{Osterbrock} \& {Pogge}(1985)}]{1985osterbrock1}
{Osterbrock}, D.~E. \& {Pogge}, R.~W. 1985, \apj, 297, 166

\bibitem[{{Padovani}(2016)}]{2016padovani1}
{Padovani}, P. 2016, \aapr, 24, 13

\bibitem[{{Pandey} \& {Sarkar}(2017)}]{2017pandey1}
{Pandey}, B. \& {Sarkar}, S. 2017, \mnras, 467, L6

\bibitem[{{Park} \& {Choi}(2009)}]{2009park1}
{Park}, C. \& {Choi}, Y.-Y. 2009, \apj, 691, 1828

\bibitem[{{Peterson} {et~al.}(2000){Peterson}, {McHardy}, {Wilkes}, {Berlind},
  {Bertram}, {Calkins}, {Collier}, {Huchra}, {Mathur}, {Papadakis}, {Peters},
  {Pogge}, {Romano}, {Tokarz}, {Uttley}, {Vestergaard}, \&
  {Wagner}}]{2000peterson1}
{Peterson}, B.~M., {McHardy}, I.~M., {Wilkes}, B.~J., {et~al.} 2000, \apj, 542,
  161

\bibitem[{{Poudel} {et~al.}(2017){Poudel}, {Hein{\"a}m{\"a}ki}, {Tempel},
  {Einasto}, {Lietzen}, \& {Nurmi}}]{2017poudel1}
{Poudel}, A., {Hein{\"a}m{\"a}ki}, P., {Tempel}, E., {et~al.} 2017, \aap, 597,
  A86

\bibitem[{{Povi{\'c}} {et~al.}(2012){Povi{\'c}}, {S{\'a}nchez-Portal},
  {P{\'e}rez Garc{\'{\i}}a}, {Bongiovanni}, {Cepa}, {Huertas-Company},
  {Lara-L{\'o}pez}, {Fern{\'a}ndez Lorenzo}, {Ederoclite}, {Alfaro},
  {Casta{\~n}eda}, {Gallego}, {Gonz{\'a}lez-Serrano}, \&
  {Gonz{\'a}lez}}]{2012povic1}
{Povi{\'c}}, M., {S{\'a}nchez-Portal}, M., {P{\'e}rez Garc{\'{\i}}a}, A.~M.,
  {et~al.} 2012, \aap, 541, A118

\bibitem[{{Rakshit} {et~al.}(2017){Rakshit}, {Stalin}, {Chand}, \&
  {Zhang}}]{2017rakshit1}
{Rakshit}, S., {Stalin}, C.~S., {Chand}, H., \& {Zhang}, X.-G. 2017, \apjs,
  229, 39

\bibitem[{{Richards} \& {Lister}(2015)}]{2015richards1}
{Richards}, J.~L. \& {Lister}, M.~L. 2015, \apjl, 800, L8

\bibitem[{{Richards} {et~al.}(2015){Richards}, {Lister}, {Savolainen}, {Homan},
  {Kadler}, {Hovatta}, {Readhead}, {Arshakian}, \&
  {Chavushyan}}]{2015richards2}
{Richards}, J.~L., {Lister}, M.~L., {Savolainen}, T., {et~al.} 2015, in IAU
  Symposium, Vol. 313, Extragalactic Jets from Every Angle, ed. F.~{Massaro},
  C.~C. {Cheung}, E.~{Lopez}, \& A.~{Siemiginowska}, 139--142

\bibitem[{{Ryan} {et~al.}(2007){Ryan}, {De Robertis}, {Virani}, {Laor}, \&
  {Dawson}}]{2007ryan1}
{Ryan}, C.~J., {De Robertis}, M.~M., {Virani}, S., {Laor}, A., \& {Dawson},
  P.~C. 2007, \apj, 654, 799

\bibitem[{{Sani} {et~al.}(2010){Sani}, {Lutz}, {Risaliti}, {Netzer}, {Gallo},
  {Trakhtenbrot}, {Sturm}, \& {Boller}}]{2010sani1}
{Sani}, E., {Lutz}, D., {Risaliti}, G., {et~al.} 2010, \mnras, 403, 1246

\bibitem[{{Schmitt}(2001)}]{2001schmitt2}
{Schmitt}, H.~R. 2001, \aj, 122, 2243

\bibitem[{{Schmitt} {et~al.}(2001){Schmitt}, {Antonucci}, {Ulvestad}, {Kinney},
  {Clarke}, \& {Pringle}}]{2001schmitt1}
{Schmitt}, H.~R., {Antonucci}, R.~R.~J., {Ulvestad}, J.~S., {et~al.} 2001,
  \apj, 555, 663

\bibitem[{{Shen} \& {Ho}(2014)}]{2014shen1}
{Shen}, Y. \& {Ho}, L.~C. 2014, \nat, 513, 210

\bibitem[{{Shen} {et~al.}(2013){Shen}, {McBride}, {White}, {Zheng}, {Myers},
  {Guo}, {Kirkpatrick}, {Padmanabhan}, {Parejko}, {Ross}, {Schlegel},
  {Schneider}, {Streblyanska}, {Swanson}, {Zehavi}, {Pan}, {Bizyaev},
  {Brewington}, {Ebelke}, {Malanushenko}, {Malanushenko}, {Oravetz}, {Simmons},
  \& {Snedden}}]{2013shen1}
{Shen}, Y., {McBride}, C.~K., {White}, M., {et~al.} 2013, \apj, 778, 98

\bibitem[{{Simpson}(1998)}]{1998simpson1}
{Simpson}, C. 1998, \mnras, 297, L39

\bibitem[{{Spergel} {et~al.}(2007){Spergel}, {Bean}, {Dor{\'e}}, {Nolta},
  {Bennett}, {Dunkley}, {Hinshaw}, {Jarosik}, {Komatsu}, {Page}, {Peiris},
  {Verde}, {Halpern}, {Hill}, {Kogut}, {Limon}, {Meyer}, {Odegard}, {Tucker},
  {Weiland}, {Wollack}, \& {Wright}}]{2007spergel1}
{Spergel}, D.~N., {Bean}, R., {Dor{\'e}}, O., {et~al.} 2007, \apjs, 170, 377

\bibitem[{{Steinhauser} {et~al.}(2016){Steinhauser}, {Schindler}, \&
  {Springel}}]{2016steinhauser1}
{Steinhauser}, D., {Schindler}, S., \& {Springel}, V. 2016, \aap, 591, A51

\bibitem[{{Storchi-Bergmann}(2008)}]{2008storchi-bergmann1}
{Storchi-Bergmann}, T. 2008, in Revista Mexicana de Astronomia y Astrofisica
  Conference Series, Vol.~32, Revista Mexicana de Astronomia y Astrofisica
  Conference Series, 139--146

\bibitem[{{Sulentic} {et~al.}(2007){Sulentic}, {Bachev}, {Marziani}, {Negrete},
  \& {Dultzin}}]{2007sulentic1}
{Sulentic}, J.~W., {Bachev}, R., {Marziani}, P., {Negrete}, C.~A., \&
  {Dultzin}, D. 2007, \apj, 666, 757

\bibitem[{{Taniguchi}(1999)}]{1999taniguchi1}
{Taniguchi}, Y. 1999, \apj, 524, 65

\bibitem[{{Tempel} {et~al.}(2014{\natexlab{a}}){Tempel}, {Stoica},
  {Mart{\'{\i}}nez}, {Liivam{\"a}gi}, {Castellan}, \& {Saar}}]{2014tempel1}
{Tempel}, E., {Stoica}, R.~S., {Mart{\'{\i}}nez}, V.~J., {et~al.}
  2014{\natexlab{a}}, \mnras, 438, 3465

\bibitem[{{Tempel} {et~al.}(2014{\natexlab{b}}){Tempel}, {Tamm}, {Gramann},
  {Tuvikene}, {Liivam{\"a}gi}, {Suhhonenko}, {Kipper}, {Einasto}, \&
  {Saar}}]{2014A&A...566A...1T}
{Tempel}, E., {Tamm}, A., {Gramann}, M., {et~al.} 2014{\natexlab{b}}, \aap,
  566, A1

\bibitem[{{Tran}(2001)}]{2001tran1}
{Tran}, H.~D. 2001, \apjl, 554, L19

\bibitem[{{Tran}(2003)}]{2003tran1}
{Tran}, H.~D. 2003, \apj, 583, 632

\bibitem[{{Treister} {et~al.}(2012){Treister}, {Schawinski}, {Urry}, \&
  {Simmons}}]{2012treister1}
{Treister}, E., {Schawinski}, K., {Urry}, C.~M., \& {Simmons}, B.~D. 2012,
  \apjl, 758, L39

\bibitem[{{Urrutia} {et~al.}(2008){Urrutia}, {Lacy}, \&
  {Becker}}]{2008urrutia1}
{Urrutia}, T., {Lacy}, M., \& {Becker}, R.~H. 2008, \apj, 674, 80

\bibitem[{{van de Ven} \& {Fathi}(2010)}]{2010vandeven1}
{van de Ven}, G. \& {Fathi}, K. 2010, \apj, 723, 767

\bibitem[{{Veron-Cetty} \& {Veron}(2003)}]{2003veroncetty1}
{Veron-Cetty}, M.~P. \& {Veron}, P. 2003, VizieR Online Data Catalog, 7235, 0

\bibitem[{{Villforth} {et~al.}(2014){Villforth}, {Hamann}, {Rosario},
  {Santini}, {McGrath}, {van der Wel}, {Chang}, {Guo}, {Dahlen}, {Bell},
  {Conselice}, {Croton}, {Dekel}, {Faber}, {Grogin}, {Hamilton}, {Hopkins},
  {Juneau}, {Kartaltepe}, {Kocevski}, {Koekemoer}, {Koo}, {Lotz}, {McIntosh},
  {Mozena}, {Somerville}, \& {Wild}}]{2014villforth1}
{Villforth}, C., {Hamann}, F., {Rosario}, D.~J., {et~al.} 2014, \mnras, 439,
  3342

\bibitem[{{Whalen} {et~al.}(2006){Whalen}, {Laurent-Muehleisen}, {Moran}, \&
  {Becker}}]{2006whalen1}
{Whalen}, D.~J., {Laurent-Muehleisen}, S.~A., {Moran}, E.~C., \& {Becker},
  R.~H. 2006, \aj, 131, 1948

\bibitem[{{Wolf}(2005)}]{2005wolf1}
{Wolf}, C. 2005, \memsai, 76, 21

\bibitem[{{Woo} {et~al.}(2015){Woo}, {Yoon}, {Park}, {Park}, \&
  {Kim}}]{2015woo1}
{Woo}, J.-H., {Yoon}, Y., {Park}, S., {Park}, D., \& {Kim}, S.~C. 2015, \apj,
  801, 38

\bibitem[{{Wu} {et~al.}(2004){Wu}, {Wang}, {Kong}, {Liu}, \& {Han}}]{2004wu1}
{Wu}, X.-B., {Wang}, R., {Kong}, M.~Z., {Liu}, F.~K., \& {Han}, J.~L. 2004,
  \aap, 424, 793

\bibitem[{{Wu} {et~al.}(2011){Wu}, {Zhang}, {Liang}, {Zhang}, \&
  {Zhao}}]{2011wu1}
{Wu}, Y.-Z., {Zhang}, E.-P., {Liang}, Y.-C., {Zhang}, C.-M., \& {Zhao}, Y.-H.
  2011, \apj, 730, 121

\bibitem[{{Xu} {et~al.}(2012){Xu}, {Komossa}, {Zhou}, {Lu}, {Li}, {Grupe},
  {Wang}, \& {Yuan}}]{2012xu1}
{Xu}, D., {Komossa}, S., {Zhou}, H., {et~al.} 2012, \aj, 143, 83

\bibitem[{{York} {et~al.}(2000){York}, {Adelman}, {Anderson}, {Anderson},
  {Annis}, {Bahcall}, {Bakken}, {Barkhouser}, {Bastian}, {Berman}, {Boroski},
  {Bracker}, {Briegel}, {Briggs}, {Brinkmann}, {Brunner}, {Burles}, {Carey},
  {Carr}, {Castander}, {Chen}, {Colestock}, {Connolly}, {Crocker}, {Csabai},
  {Czarapata}, {Davis}, {Doi}, {Dombeck}, {Eisenstein}, {Ellman}, {Elms},
  {Evans}, {Fan}, {Federwitz}, {Fiscelli}, {Friedman}, {Frieman}, {Fukugita},
  {Gillespie}, {Gunn}, {Gurbani}, {de Haas}, {Haldeman}, {Harris}, {Hayes},
  {Heckman}, {Hennessy}, {Hindsley}, {Holm}, {Holmgren}, {Huang}, {Hull},
  {Husby}, {Ichikawa}, {Ichikawa}, {Ivezi{\'c}}, {Kent}, {Kim}, {Kinney},
  {Klaene}, {Kleinman}, {Kleinman}, {Knapp}, {Korienek}, {Kron}, {Kunszt},
  {Lamb}, {Lee}, {Leger}, {Limmongkol}, {Lindenmeyer}, {Long}, {Loomis},
  {Loveday}, {Lucinio}, {Lupton}, {MacKinnon}, {Mannery}, {Mantsch}, {Margon},
  {McGehee}, {McKay}, {Meiksin}, {Merelli}, {Monet}, {Munn}, {Narayanan},
  {Nash}, {Neilsen}, {Neswold}, {Newberg}, {Nichol}, {Nicinski}, {Nonino},
  {Okada}, {Okamura}, {Ostriker}, {Owen}, {Pauls}, {Peoples}, {Peterson},
  {Petravick}, {Pier}, {Pope}, {Pordes}, {Prosapio}, {Rechenmacher}, {Quinn},
  {Richards}, {Richmond}, {Rivetta}, {Rockosi}, {Ruthmansdorfer}, {Sandford},
  {Schlegel}, {Schneider}, {Sekiguchi}, {Sergey}, {Shimasaku}, {Siegmund},
  {Smee}, {Smith}, {Snedden}, {Stone}, {Stoughton}, {Strauss}, {Stubbs},
  {SubbaRao}, {Szalay}, {Szapudi}, {Szokoly}, {Thakar}, {Tremonti}, {Tucker},
  {Uomoto}, {Vanden Berk}, {Vogeley}, {Waddell}, {Wang}, {Watanabe},
  {Weinberg}, {Yanny}, {Yasuda}, \& {SDSS Collaboration}}]{2000york1}
{York}, D.~G., {Adelman}, J., {Anderson}, Jr., J.~E., {et~al.} 2000, \aj, 120,
  1579

\bibitem[{{Yu} {et~al.}(2013){Yu}, {Huang}, {Hwang}, \& {Ohyama}}]{2013yu1}
{Yu}, P.-C., {Huang}, K.-Y., {Hwang}, C.-Y., \& {Ohyama}, Y. 2013, \apj, 768,
  30

\bibitem[{{Yu} \& {Hwang}(2011)}]{2011yu1}
{Yu}, P.-C. \& {Hwang}, C.-Y. 2011, \aj, 142, 14

\bibitem[{{Yuan} {et~al.}(2008){Yuan}, {Zhou}, {Komossa}, {Dong}, {Wang}, {Lu},
  \& {Bai}}]{2008yuan1}
{Yuan}, W., {Zhou}, H.~Y., {Komossa}, S., {et~al.} 2008, \apj, 685, 801

\bibitem[{{Zhang} \& {Wang}(2006)}]{2006zhang1}
{Zhang}, E.-P. \& {Wang}, J.-M. 2006, \apj, 653, 137

\bibitem[{{Zhou} {et~al.}(2006){Zhou}, {Wang}, {Yuan}, {Lu}, {Dong}, {Wang}, \&
  {Lu}}]{2006zhou1}
{Zhou}, H., {Wang}, T., {Yuan}, W., {et~al.} 2006, \apjs, 166, 128

\bibitem[{{Zhou} {et~al.}(2007){Zhou}, {Wang}, {Yuan}, {Shan}, {Komossa}, {Lu},
  {Liu}, {Xu}, {Bai}, \& {Jiang}}]{2007zhou1}
{Zhou}, H., {Wang}, T., {Yuan}, W., {et~al.} 2007, \apjl, 658, L13

\end{thebibliography}

\begin{appendix}

\section{The LDF}
\label{densityfield}
     
We calculated the luminosities of the galaxies using $k$-corrections from the \texttt{kcorrect} algorithm \citep{Blanton2007}. To set the mean luminosity the same through the whole 
redshift range, we calculated a distance-dependent weighting factor $W_L(d)$ for the luminosities of galaxies. We calculated the mean luminosity density as a function of distance,
smoothed it, and used this as a weight.
We then calculated the weighted luminosity for each galaxy as
    \begin{equation}
      L_{\mathrm{gal,w}}=W_L(d)L_{\mathrm{gal}},
    \end{equation}

where $L_{\mathrm{gal}}=L_{\odot}10^{0.4(h^{-1}M_\odot-M)}$ is the observed luminosity of a galaxy with the absolute magnitude $M,$ and $M_\odot$ is the absolute magnitude of the Sun.

The coordinates of galaxies were transformed into Cartesian coordinates defined as

\begin{align}
  x&=-d\sin{\lambda}\\
  y&=d\cos{\lambda}\cos{\eta}\\
  z&=d\cos{\lambda}\sin{\eta},
\end{align}

where $d$ is the distance of the galaxy and $\eta$ and $\lambda$ are the SDSS angular coordinates. We then defined a 3\,$h^{-1}$Mpc grid in these coordinates for calculating the LDF.

The luminosity-density value $l_i$ at grid point $\mathbf{r}_i$ is calculated by a kernel sum

\begin{equation} l_i=\frac{1}{a^3}\sum_{\mathrm{gal}}K^{(3)}\left(\frac{\mathbf{r}_{\mathrm{gal}}-\mathbf{r}_i}{a} \right)L_{\mathrm{gal,w}},
\end{equation}

where $a$ is the kernel scale. As the kernel we used the $B_3$ spline function

\begin{equation}
B_3(x)=\frac{\mid x-2\mid^3-4\mid x-1\mid^3+6\mid x\mid^3-4\mid x+1\mid^3+\mid x+2\mid^3}{12}
,\end{equation}

with 8\,$h^{-1}$Mpc (BOSS) or 16\,$h^{-1}$Mpc (LRG) smoothing scale. As the final step in constructing the LDF, 
we normalised the density field by converting the densities into units of mean density as

\begin{equation}
D_i=\frac{l_i}{l_{\mathrm{mean}}}
,\end{equation}

where $l_{\mathrm{mean}}$ is the average over all density grid points.

\end{appendix}

\end{document}